\documentclass[aps,prl,twocolumn,superscriptaddress,amsmath,amssymb,floatfix]{revtex4}
\usepackage{color}
\usepackage{graphicx}
\usepackage{bm}

\begin{document}


\title{Charge transfer in ultracold gases via Feshbach resonances}


\author{Marko Gacesa}
\affiliation{NASA Ames Research Center, Moffett Field, CA 94035, USA}
\affiliation{Department of Physics, University of Connecticut, Storrs, CT 06269-3046, USA}

\author{Robin C\^ot\'e}
\affiliation{Department of Physics, University of Connecticut, Storrs, CT 06269-3046, USA}

\date{\today}

\begin{abstract}
We investigate the prospects of using magnetic Feshbach resonance to control charge exchange in ultracold collisions of heteroisotopic combinations of atoms and ions of the same element. The proposed treatment, readily applicable to alkali or alkaline-earth metals, is illustrated on cold collisions of $^9$Be$^{+}$ and $^{10}$Be.
Feshbach resonances are characterized by quantum scattering calculations in a 
coupled-channel formalism that includes non-Born-Oppenheimer terms originating 
from the nuclear kinetic operator. Near a resonance predicted at 322 G, 
we find the charge exchange rate coefficient to rise from practically 
zero to values greater than $10^{-12}$ cm$^3$/s.
Our results suggest controllable charge exchange processes between different isotopes of suitable atom-ion 
pairs, with potential applications to quantum systems engineered to study charge diffusion in trapped cold atom-ion mixtures and emulate many-body physics.
\end{abstract}


\maketitle


Since the original studies of ultracold atom-ion scattering \cite{2000PhRvA..62a2709C,2000PhRvL..85.5316C,2003PhRvA..67d2705M}, advances in experimental techniques for direct manipulation of small ensembles composed of ultracold atoms and ions have opened new research venues \cite{2014arXiv1401.1699W,2014ConPh..55...33H,2016AAMOP..65...67C}. 
Production and trapping of cold ions below the critical mass ratio 
is making possible direct studies of atom-ion collisional dynamics and chemistry at ultracold temperatures \cite{2014NatCo...5E5587H,2015JPhCS.635a2012E,2016NatCo...712448S,2016PhRvL.116w3003H,2016JPhB...49j5202R}, surpassing the temperature limitations of the experiments in dual overlapping traps of ions and neutrals  \cite{2005JMOp...52.2253S,2009PhRvL.102v3201G,2010PhRvL.105n3001T,2010Natur.464..388Z,2011PhRvL.107x3201R,2012PhRvA..86f3419S,2012NatCo...3E1126R}. Of particular interest are applications of such hybrid systems to atom-ion interactions and chemistry \cite{2005PhRvL..95r3002B,2008PhRvA..78d2709R,2012NatPh...8..649R,2015PhRvA..91c2709H,2016PhRvA..94c0701K}, precision measurements \cite{2014Natur.506...71B}, many-body physics \cite{2000PhRvL..85.5316C,2002PhRvL..89i3001C,2010PhRvL.105m3202S,2012PhRvL.109h0402G,2013PhRvL.111h0501B,2014PhRvA..90c3601S,2016PhRvA..93f3602S,2016PhRvA..94a3420S}, and quantum information processing \cite{2010PhRvA..81a2708D,2011LaPhL...8..188W}. 
In addition, a number of theoretical proposals suggest different approaches to forming ultracold molecular ions in their ground state \cite{2013PhRvA..87e2717S,2014JPhB...47n5201M,2014JPhCS.572a2009Y,2015PhRvA..91d2706T,2015NJPh...17d5015D,2016PhRvA..94a3407G,2016arXiv160807043S,2016JPhB...49x5202S}.
Developing a solid understanding of two-body atom-ion interactions at ultra-low temperatures, and the means to control them \textit{e.g.}, using external electric, magnetic, or optical fields, is central to advancing such research and constitutes an essential step towards envisioning new experiments and developing many-body theoretical descriptions. 
In atom-ion mixtures, as in our studies in neutrals \cite{RbCr2005,LiNa2008,LiCs2009}, Feshbach resonances (FRs) are expected to play a key role in controlling the strength and sign of the two-body interaction, consequently altering macroscopic properties of the ensemble \cite{1976PhRvL..37.1628S,kohler:1311,chin2010feshbach}. 

First theoretical studies of FRs within a given asymptotic charge arrangement in cold atom-ion mixtures have already been undertaken \cite{2000JPhB...33.5329E,2011NJPh...13h3005I,2014PhRvA..89e2704L,2015PhRvA..92f2701T}.
Additionally, near-resonant atom-ion scattering in heteroisotopic mixtures (of H, Be, Li, Rb, and Yb) has been explored 
theoretically \cite{2000JPhB...33.5329E,2000JPhB...33R..93S,2009PhRvA..80c0703Z,2008NJPh...10c3024B,2009JPCA..11315085Z,2011PCCP...1319026Z,PhysRevLett.117.143201} where it has been shown that non-Born-Oppenheimer (non-BO) couplings originating from the nuclear kinetic operator influence charge exchange (CX) and scattering cross sections in cold mixtures. These investigations, which did not include hyperfine nor Zeeman interactions, showed that this effect is especially significant in the ultracold regime. In particular, it could strongly affect transport properties of charged particles, including their diffusion and mobility (see Ref. \cite{2000PhRvL..85.5316C}), with broader consequences for engineered atom-ion systems, \textit{e.g.}, an ion immersed in a Bose-Einstein condensate (BEC) \cite{2002PhRvL..89i3001C,2015NJPh...17e3046K,2015PhRvL.114x3003W}.
Since resonant CX between identical ion and parent atom is impossible to distinguish from elastic scattering at ultracold temperatures, our choice of near-resonant CX in heteroisotopic atom-ion pair provides a simple way to discriminate among the processes.

\begin{figure}[t]
 \centering
 \includegraphics[clip,width=\linewidth]{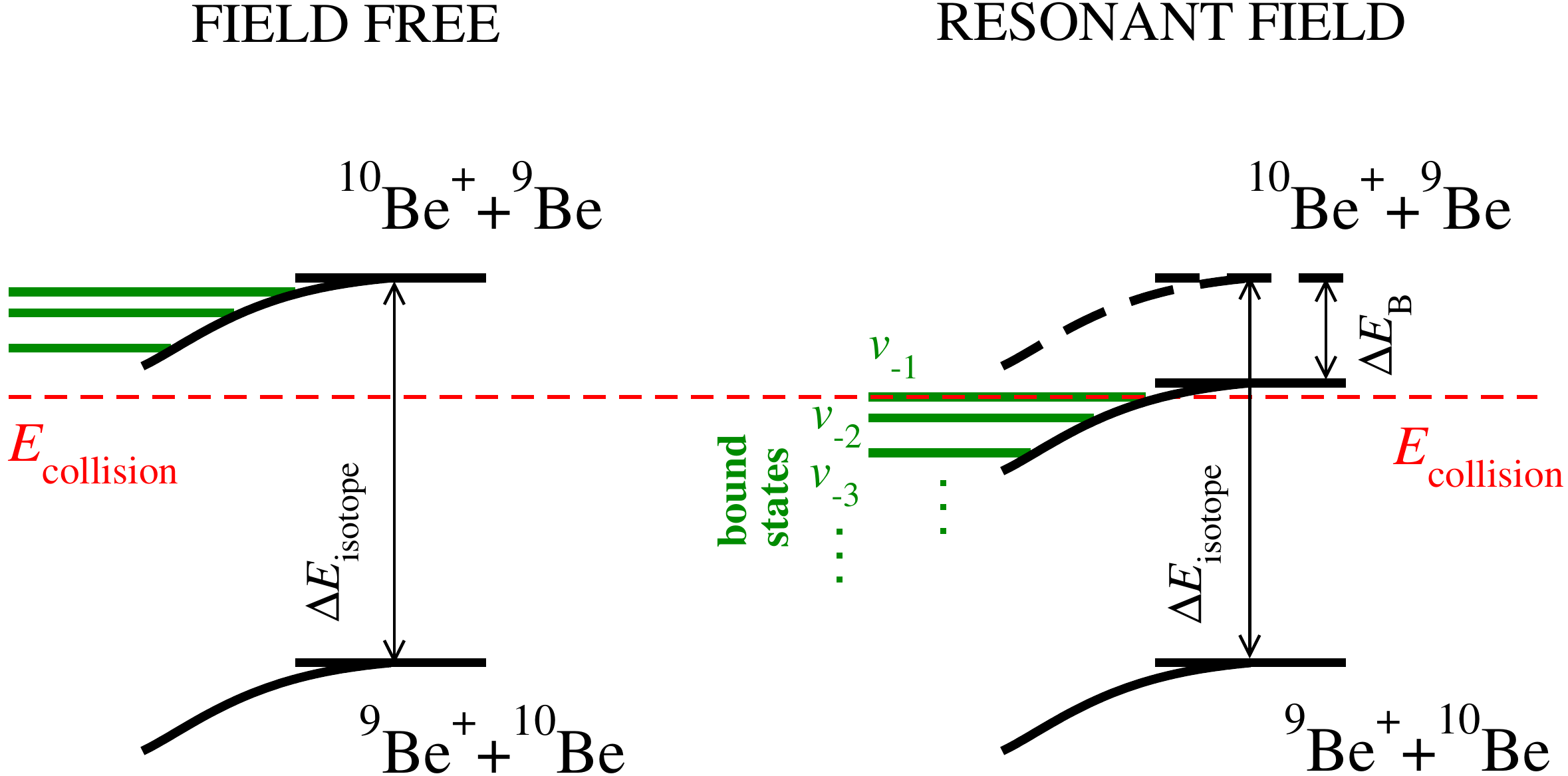}
 \caption{(Color online) Schematics of the process. \textit{Left:} Field-free scattering for a collision energy $E_\mathrm{collision}$ smaller than the asymptotic isotope shift $\Delta E_\mathrm{isotope}$. Only the lower asymptote is energetically accessible and CX to the upper asymptote does not occur. \textit{Right}: 
Magnetic field $B$ present. The upper asymptote, dressed by $B$, is shifted down by $\Delta E_B$ so 
that $E_{\rm collision}$ coincides with a closed channel in the vicinity of a bound state. At specific $B$ fields, charge exchange can be resonantly enhanced.}
 \label{fig0}
\end{figure}

In this study, we show that magnetic Feshbach resonances can be used to control the charge exchange between asymptotic charge arrangement of different isotopes for selected atom-ion mixtures (Fig. \ref{fig0}). 
The treatment below applies to systems with one valence electron ({\it e.g.}, alkali atoms, including H),
or one valence hole ({\it e.g.}, alkaline-earth and some rare earth atomic ions, like Yb$^+$), where at least one of the partners
has an hyperfine structure. We note that since both alkaline-earth
atoms and ions possess closed optical transitions, it is possible to image them separately, as opposed to alkali species for which ions cannot be optically imaged.
For this reason, we focus our attention on alkaline-earth mixtures. Without loss of generality, we select the simplest element, Be, and study the CX process in cold collision of $^9$Be + $^{10}$Be$^+$ in an external magnetic field. For this system, the molecular ion potential energy curves as well as their non-BO corrections are known \cite{2011PCCP...1319026Z} and the value of the isotope shift energy between the two lowest asymptotes suggests that Feshbach resonances will occur at experimentally attainable magnetic fields (Fig. \ref{fig0}). Moreover, the fact that $^9$Be has a non-zero nuclear spin ($I=3/2$) and $^{10}$Be has zero nuclear spin is suitable for exploring many-body charge dynamics in quantum systems.

If the kinetic energy operator in the center of nuclear mass (CNM) motion is separated out, a collision between an 
alkaline-earth atom and its ion in a magnetic field can be described by an effective Hamiltonian in the body-fixed frame as 
\cite{PhysRevA.51.4852,kohler:1311} 
\begin{equation}
 \hat{H} = \hat{T}_N + \hat{T}_e + \hat{T}_{\mathrm{mp}} + \hat{V}(\vec{r},\vec{R}) + \hat{H}_\mathrm{int} ,
 \label{eq:1}
\end{equation}
where, $\hat{T}_{\mathrm{N}}$ is the kinetic energy operator describing the relative motion of the nuclei, $\hat{T}_e$ is the electronic kinetic energy operator, $\hat{V}(\vec{r},\vec{R})$ is the electrostatic interaction operator, $\vec{r}$ are electronic coordinates in the CNM frame, and $\vec{R}$ is the vector connecting the two nuclei.
$\hat{T}_{\mathrm{mp}}$ is the mass polarization term \cite{2011PCCP...1319026Z}, defined as
\begin{equation}
 \hat{T}_\mathrm{mp} = -\frac{1}{2m} \sum_{i,j=1}^{N_e} \vec{\nabla}_i \vec{\nabla}_j,
\end{equation}
where $m = m_1 + m_2$ is the total mass of the nuclei of masses $m_1$ and $m_2$, and the summation runs over the total number of electrons $N_e$.
The operator $\hat{H}_\mathrm{int} = \hat{V}_B + \hat{V}_{\mathrm{hf}}$ describes the interactions of the particles' internal degrees of freedom, namely electronic and nuclear spins, in the external magnetic field, $\vec{B}=B \hat{z}$.
Specifically, $\hat{V}_B$ can be written as
\begin{equation}
 \hat{V}_B = 2 \mu_0 B (\hat{S}_{z_1} + \hat{S}_{z_2}) - 
       B \left( \frac{\mu_1}{I_1}\hat{I}_{z_1} + \frac{\mu_2}{I_2}\hat{I}_{z_2} \right) ,
\end{equation}
where $\mu_0$ is the Bohr magneton, $\hat{S}_{z_i}$ ($i=1,2$) is the $z$-projection electronic spin 
operator of the colliding particles, with corresponding nuclear magnetic moment $\mu_i$, nuclear spin quantum number
$I_i$, and $z$-projection nuclear spin operator $\hat{I}_{z_i}$.
The term $\hat{V}_{\mathrm{hf}}$ is given by
\begin{equation}
 \hat{V}_{\mathrm{hf}} = \sum_{i=1}^{2} \alpha_{\mathrm{hf}}^{(i)} \hat{I_i} \cdot \hat{S_i},
\end{equation}
where $\alpha_{\mathrm{hf}}^{(i)}$ is the atomic hyperfine constant ($i=1,2$).

The total wave function $\Psi(\vec{r},\vec{R})$ can be expanded as a sum of partial waves given by the total angular momentum quantum number $J$ and its projection $M$ onto the internuclear axis, where, for each $(J,M)$ we have
\begin{equation}
 \Psi^{J,M}(\vec{r},\vec{R}) = \frac{e^{iM\varphi}}{R} \sum_{\alpha, \Lambda} 
    \chi_{\alpha,\Lambda}^{J}(R) \Theta_{M,\Lambda}^{J} (\vartheta) | \alpha \rangle,
    \label{eq:totalwf}
\end{equation}
where $\Lambda$ is the projection of $\vec{L}$, the total orbital angular momentum of the electrons, onto the internuclear axis.  
The angles of the vector $\vec{R}$ in spherical coordinates are $\vartheta$ and $\varphi$, 
$\Theta_{M,\Lambda}^{J} (\vartheta)$ is the corresponding generalized spherical harmonic, $\chi_{\alpha,\Lambda}^{J}(R)$ is the radial wave function describing nuclear motion, and $|\alpha\rangle$ is an eigenstate composed of a BO electronic state and  nuclear and electronic spins of the particles. 
Alkali and alkaline-earth elements have $s$-state valence electrons (or holes)  
with no orbital angular momentum ($L_i=0$), so that $\Lambda=0$ (since $\vec{L}=\vec{L}_1+\vec{L}_2=0$) and 
the total spin is $\vec{J}=\vec{I}+\vec{S}$, where $\vec{S} = \vec{S}_1 + \vec{S}_2$ and $\vec{I} = \vec{I}_1 + \vec{I}_2$. 
In the coupled molecular basis, we can express the state $| \alpha \rangle$ as $|\alpha\rangle =  | n J M ; S M_S I M_I \rangle$, 
where $n$ is the ordinal number of the electronic state of the same symmetry, while $S$ and $I$ and the total spin and nuclear spin quantum numbers associated with $\vec{S}$ and $\vec{I}$, respectively, with their projections onto the internuclear axis given by $M_S$ and $M_I$. The eigenvalues of the non-relativistic Born-Oppenheimer electronic Hamiltonian, $\hat{H}_{\mathrm{el}} = \hat{T}_e + \hat{V}(\vec{r},\vec{R})$, for the state $|\alpha \rangle$ are denoted as $\varepsilon_{\alpha}$.

Since we are primarily interested in dynamics at ultracold conditions, where only the lowest asymptote is energetically accessible, we restrict our analysis to the two energetically lowest electronic states. This approximation is valid for systems whose electronic structure is similar to that of a dimer composed of an alkaline-earth atom and alkaline-earth ion, where a single valent electron (or hole) can participate in the CX process. 
In addition, we assume \textit{s}-wave collisions ($J=0, \Lambda = 0, M=0$) and neglect higher partial waves. 
Consequently, we label the two electronic states as $n=1$ and 2, respectively. Due to the internal spin degrees of freedom, for each of the electronic states there will exist a total of $N_\mathrm{hf}=(2 I_1+1)(2 S_1+1)(2 I_2+1)(2 S_2+1)$ hyperfine states. 

We insert Eq. (\ref{eq:totalwf}) in the time-independent Schr\"odinger equation and integrate out the electronic coordinates to obtain a system of coupled-channel equations for the radial wave functions $\chi_\beta(R)$ at the interaction energy $E$
(in atomic units): 
\begin{eqnarray}
\left(-\frac{1}{2\mu} \frac{d^2}{dR^2} + \varepsilon_\beta - E\right) \chi_\beta + 
  \sum_\alpha \varepsilon_{\beta \alpha}^\mathrm{mp} \chi_\alpha = \nonumber \\ 
\frac{1}{\mu} \sum_{\alpha \neq \beta} \langle \beta | \frac{\partial}{\partial R} | \alpha \rangle \frac{d \chi_\alpha}{d R} +
  \frac{1}{2 \mu} \sum_\alpha \langle \beta | \frac{\partial^2}{\partial R^2} | \alpha \rangle \chi_\alpha + \nonumber \\ 
  \sum_\alpha \langle \beta | \hat{H}_\mathrm{int} | \alpha \rangle \chi_\alpha  \, ,
\end{eqnarray}
where $\mu = (m_9 m_{10})/(m_9+m_{10})$ is the reduced mass. We used $m_9=9.0121821$ and $m_{10}=10.013534$ amu.
The system can be expressed in a matrix form:
\begin{equation}
 \left[ \mathbf{I} \frac{d^2}{dR^2} + 2 \mathbf{F} \frac{d}{dR} + \mathbf{k}^2 - 
    2\mu (\mathbf{V} - \mathbf{Z}) \right] \boldsymbol{\chi} = 0
    \label{eq:matrixform1}
\end{equation}
where \textbf{I} is the identity matrix of dimension $2N_\mathrm{hf}$, \textbf{k}$^2$ is a diagonal matrix of electronic states' threshold energies $k^2_j$, that satisfy $k^2_{\beta} - k^2_{\alpha} = 2\mu \Delta E$, where $\Delta E$ is the threshold energy.
The matrix \textbf{F} is given by
\begin{equation}
 F_{\alpha \beta} = -F_{\beta \alpha} = \langle \alpha | \frac{\partial}{\partial R} | \beta \rangle .
 \label{eq:F}
 \end{equation}
The matrix \textbf{V} originates from the radial and angular part of the nuclear kinetic energy operator, including 
the mass polarization. Its elements are 
\begin{equation}
  V_{\alpha \beta} = \delta_{\alpha \beta} \varepsilon_\alpha + \varepsilon_{\alpha \beta}^\mathrm{mp} + 
     \frac{1}{2\mu} \langle \alpha | \frac{\partial^2}{\partial R^2} | \beta \rangle .
\end{equation}
The matrix \textbf{Z} includes the hyperfine and Zeeman interactions given by the last two terms in Eq. (\ref{eq:1}):
\begin{equation}
  Z_{\alpha \beta} = \langle {\alpha} | (\hat{V}_{\mathrm{hf}} + \hat{V}_B) | \beta \rangle .
  \label{eq:Z}
\end{equation}
It is convenient to define a hermitian matrix $\tilde{\textbf{V}}$ with 
\begin{equation}
\tilde{V}_{\alpha \beta} = V_{\alpha \beta} - \frac{1}{2\mu} \frac{d}{dR} F_{\alpha \beta} ,
\end{equation}
and rewrite Eq. (\ref{eq:matrixform1}) so that the proper asymptotic scattering boundary conditions are restored in the atomic representation. The coupled-channel matrix equation is obtained by adding and subtracting two Eqs. (\ref{eq:matrixform1}) and substituting $\tilde{\textbf{V}}$:
\begin{equation}
 \left[ \mathbf{I} \frac{d^2}{dR^2} + 2 \mathbf{F} \frac{d}{dR} + \mathbf{k}^2 + 2\mu \mathbf{Z} - 
    2\mu \mathbf{C} \right] \boldsymbol{\tilde{\chi}} = 0 ,
    \label{eq:matrixform2}
\end{equation}
where \textbf{C} is defined by
\begin{eqnarray}
  C_{\alpha \alpha} & = & \frac{1}{2}(\tilde{V}_{\alpha \alpha} + \tilde{V}_{\beta \beta}) + \tilde{V}_{\alpha \beta} \nonumber \\
  C_{\beta \beta} & = & \frac{1}{2}(\tilde{V}_{\alpha \alpha} + \tilde{V}_{\beta \beta}) - \tilde{V}_{\alpha \beta}  \nonumber \\
  C_{\alpha \beta} & = & \frac{1}{2}(\tilde{V}_{\alpha \alpha} - \tilde{V}_{\beta \beta}) - \frac{d}{dR} F_{\alpha \beta}  \nonumber \\
  C_{\beta \alpha} & = & \frac{1}{2}(\tilde{V}_{\alpha \alpha} - \tilde{V}_{\beta \beta}) + \frac{d}{dR} F_{\alpha \beta} ,
  \label{eq:matrixC}
\end{eqnarray}
with $\alpha=1 \dotsc N_\mathrm{hf}$, and $\beta=(N_\mathrm{hf}+1) \dotsc 2 N_\mathrm{hf}$.
%

\begin{figure}[t]
 \centering
 \includegraphics[clip,width=\linewidth]{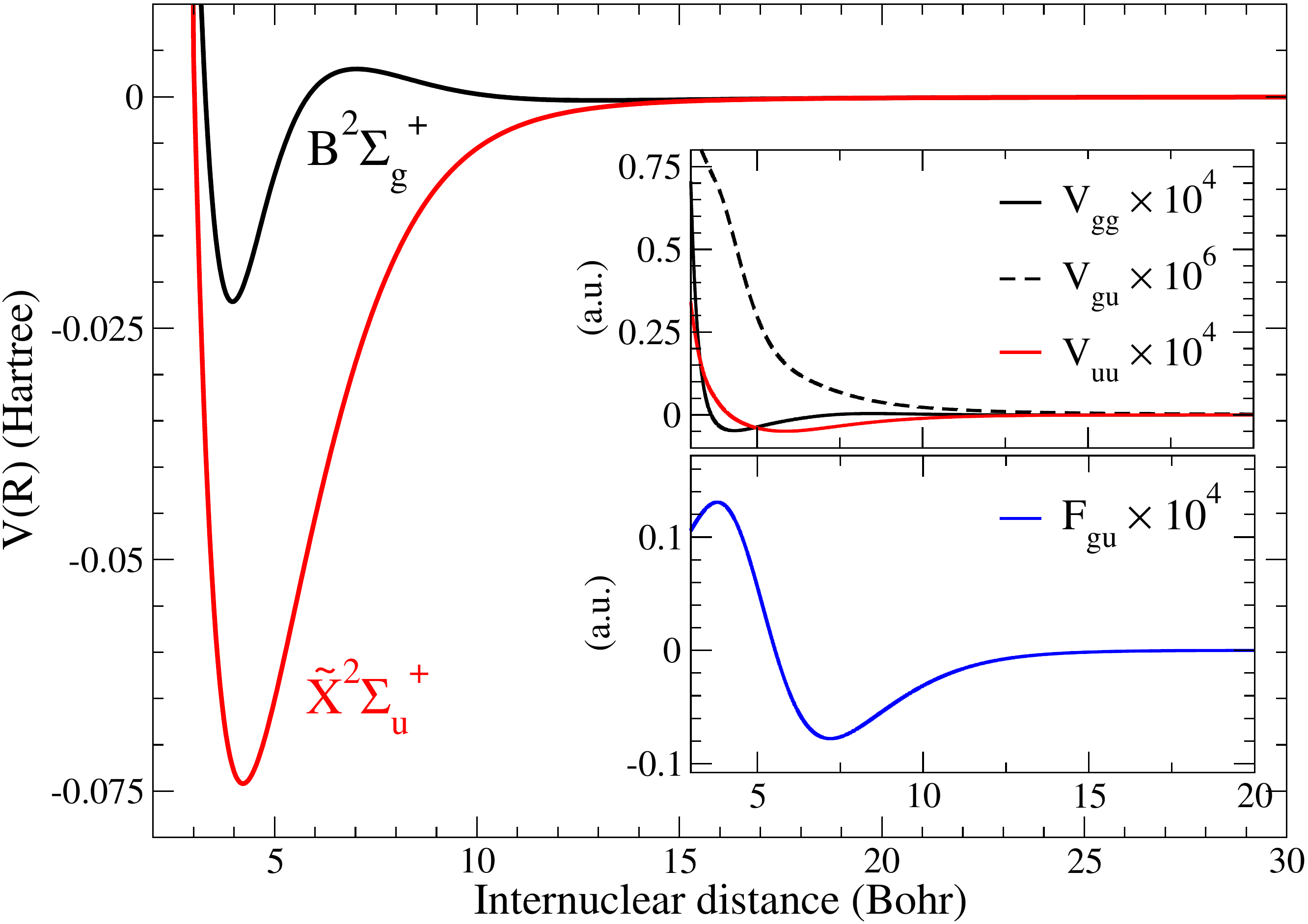}
 \caption{(Color online) Electronic potentials and non-Born-Oppenheimer couplings for ($^9$Be$^{10}$Be)$^+$ molecular ion.}
 \label{fig:potentials}
\end{figure}

Let us take a closer look at the matrix elements in Eqs. (\ref{eq:F}-\ref{eq:matrixform2}). The matrix \textbf{F} is off-diagonal and contains first derivative couplings between the two electronic states. The diagonal elements of the matrix \textbf{V} introduce non-BO corrections to $\varepsilon_\alpha$, while its off-diagonal elements couple the two electronic states and asymptotically separate their eigenvalues to the correct limits: $\pm \Delta E / 2$. See Ref. \cite{PhysRevA.59.1309} for a detailed discussion.
The matrix \textbf{Z} contains couplings between electronic and nuclear spins and the external magnetic field $B$. For the non-zero value $B$, it breaks the degeneracy of hyperfine states' manifolds and splits them into Zeeman sublevels. 
This phenomenon is responsible for the existence of magnetic Feshbach resonances in diatomic collisions of neutrals \cite{chin2010feshbach}. 
In case of different nuclear spins and different charge arrangements (\textit{e.g.}, $^9$Be$^{10}$Be$^+$ vs $^{10}$Be$^9$Be$^+$), in a non-zero $B$ field, the matrix \textbf{Z} splits each hyperfine state into Zeeman sublevels that will be coupled if permitted by the symmetry (for $B>0$ the total projection quantum number $M_F = M_S + M_I$ is conserved and the nuclear spins of two particles are equal, while the total electronic spin $S>0$).
However, the nuclear kinetic operator (matrices \textbf{F} and \textbf{V}) will couple individual hyperfine channels as long as the total projection $M_F$ is conserved, giving rise to Feshbach resonances between the states of different asymptotic charge arrangement and leading to magnetically controllable charge exchange. 
Note that asymptotic couplings in coupled-channel systems in the context of low-energy atomic collisions were discussed in detail by Grosser, Menzel, and Belyeav \cite{PhysRevA.59.1309}, whose work justifies our approach and clarifies several points related to a choice of coordinate systems that are outside of the scope of this article. 


\begin{figure}[t]
 \centering
 \includegraphics[clip,width=\linewidth]{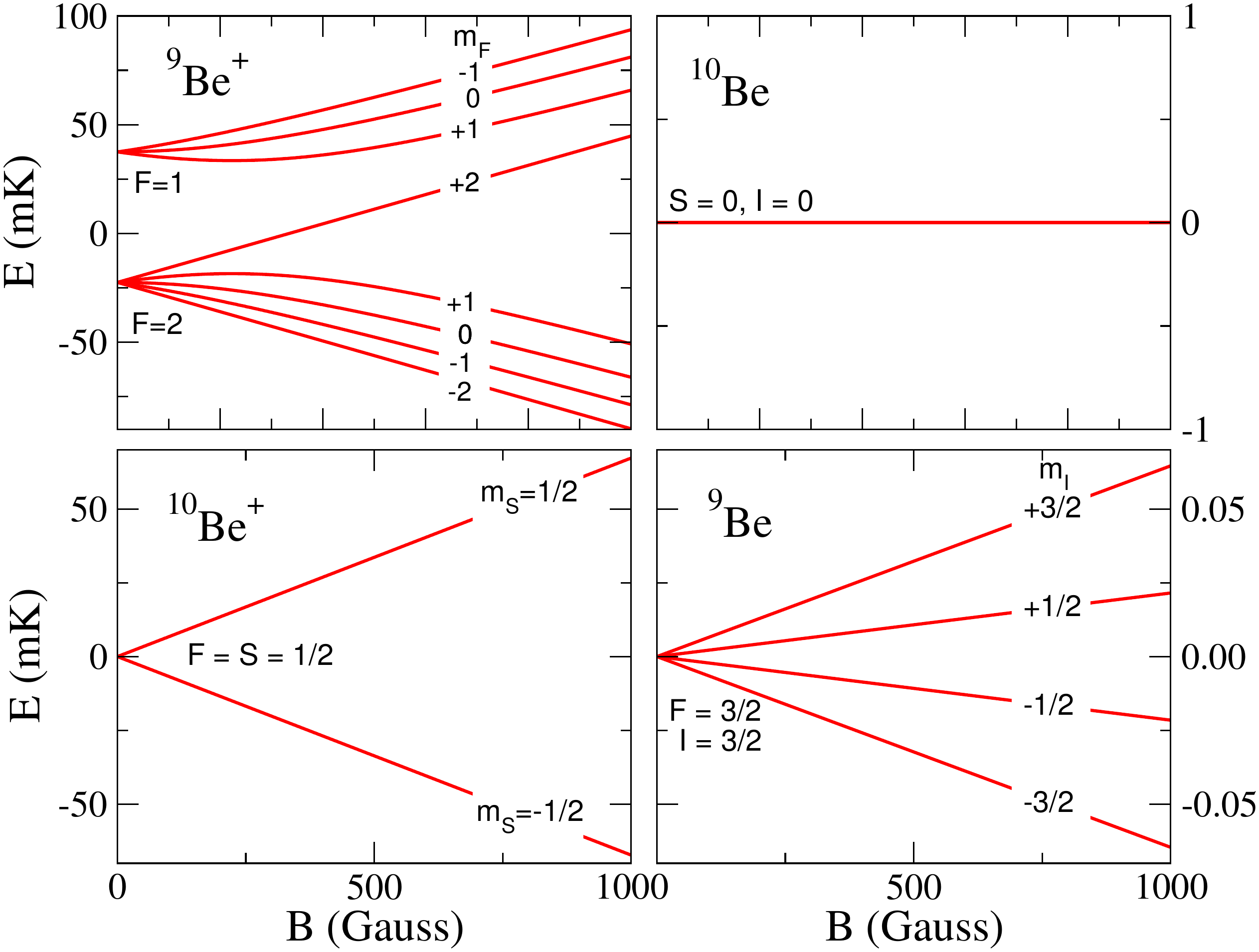}
 \caption{(Color online) Zeeman splittings of hyperfine states for $^9$Be, $^9$Be$^+$, $^{10}$Be and $^{10}$Be$^+$.}
 \label{fig:zeeman}
\end{figure}

To illustrate the effects of controlling charge transfer via Feshbach resonances we apply our model to $^9$Be$^+$+$^{10}$Be in the external magnetic field $\vec{B}=B \hat{z}$. The electronic potential energy curves (PECs) $\tilde{X}^2\Sigma_{u}^{+}$ and $B^2\Sigma_{g}^+$ relevant for this work were studied previously \cite{2011PCCP...1319026Z,Banerjee2010208,2013MolPh.111.2292L}, and the effects of the nuclear kinetic operator and isotope shift were calculated \cite{2011PCCP...1319026Z}. 
We adopted the PECs from Ref. \cite{Banerjee2010208}, the non-BO couplings from Ref. \cite{2011PCCP...1319026Z}, and parametrized the electronic exchange as $V_\mathrm{exch}(R)=A R^{\alpha} e^{-\beta R} \left( 1 + B/R \right)$, where $A=0.6316$, $B=-2.72527$, $\alpha = 1.416097$, $\beta = 0.827781$, all in atomic units. We have implicitly assumed the uncertainty of the \textit{ab-initio} methods in the exchange region of up to 10\% and selected the values that fit Refs. \cite{2011PCCP...1319026Z} and \cite{Banerjee2010208}. The curves were smoothly connected to the long-range form $V_\mathrm{lr}(R) = \pm V_\mathrm{exch}(R) - C_4 R^{-4} - C_6 R^{-6}$ at about $R=20$ Bohr, where $C_4 = 19.06$ and $C_6 = 274.2$ a.u., and the positive (negative) value of $V_\mathrm{exch}$ corresponds to $B^2\Sigma_{g}^+$ $(\tilde{X}^2\Sigma_{u}^{+})$, respectively.
The final PECs and couplings are given in Fig. \ref{fig:potentials}. 

For a non-zero magnetic field, the hyperfine states will split into Zeeman sublevels according to the nuclear spins of the Be isotopes: $I=3/2$ for $^9$Be and $I=0$ for $^{10}$Be (Fig. \ref{fig:zeeman}). Note that the electronic spin is $S=0$ for atomic Be and $S=1/2$ for Be$^+$ ion. Consequently, there will be $N_\mathrm{hf}=8$ channels for each charge arrangement $\lambda$ ($8 \times 1$ for $^9$Be$^+$+$^{10}$Be (arrangement $\lambda$=1) and $2 \times 4$ for $^{10}$Be$^+$+$^9$Be (arrangement $\lambda$=2) for a total of 16 coupled channels. 
The channels for which the total $M_F$ is conserved will be coupled, reducing the system size depending on the choice of the entrance channel.


\begin{figure}[t]
 \centering
 \includegraphics[clip,width=\linewidth]{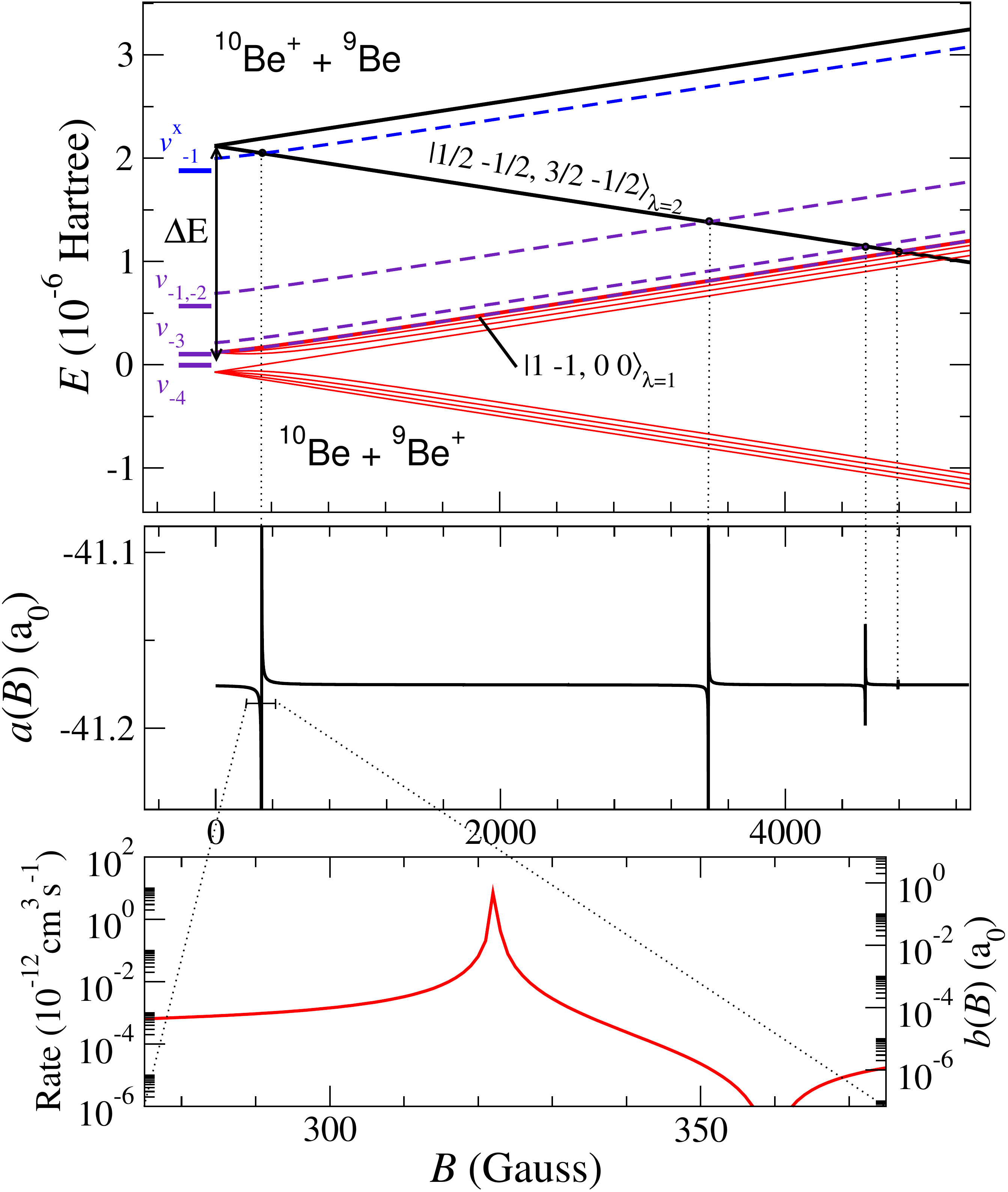}
 \caption{(Color online) 
  \textit{Top:} Asymptotic model of Feshbach-resonant bound states. Two sets of curves indicate Zeeman-split hyperfine manifolds (for $\lambda=1$ (red) and $\lambda=2$ (black)) separated by the isotope shift energy $\Delta E$. Near-dissociation bound vib. levels $v_\mathrm{-1}(\tilde{X})$ and $v_{-1},..,v_{-4}$ are shown for $\tilde{X}^2\Sigma^{+}$ (blue) and $B^2\Sigma^+$ (violet) states, respectively, while thick dashed lines (blue for $\tilde{X}^2\Sigma^{+}$ and violet for $B^2\Sigma^+$) correspond to the entrance channel $|1\rangle_{\lambda=1}$ shifted up by their binding energy. The crossings of thick curves indicate magnetic fields for which Feshbach resonances occur.
 \textit{Middle:} Real part of the scattering length $a(B)$ given as a function of magnetic field $B$ for the entrance channel $|1\rangle$. The dotted vertical lines indicate the resonant hyperfine states based on the asymptotic model.
 \textit{Bottom:} Charge exchange rate and the corresponding imaginary part $b$ of the complex scattering rate given as a function of $B$ near the resonance at 322 G.
}
 \label{fig3}
\end{figure}

To illustrate the magnetically controlled charge exchange, we select $|1\rangle = | 1 ,-1, 0, 0\rangle_{\lambda=1}$, in the $|F_1, m_{F_1}, F_2, m_{F_2}\rangle_\lambda$ basis, as the entrance channel with the total projection quantum number $M_F=-1$.
Here, for $\lambda=1$, $|F_1, m_{F_1}\rangle$ correspond to $^9$Be$^+$, $|F_2, m_{F_2}\rangle$ to $^{10}$Be and, for $\lambda=2$, $|F_1, m_{F_1}\rangle$ to $^{10}$Be$^+$, and $|F_2, m_{F_2}\rangle$ to $^9$Be.
 It couples to an open channel $|2\rangle = | 2, -1, 0, 0\rangle_{\lambda=1}$ and two closed channels in the upper hyperfine sublevel manifold: $|3\rangle = | \frac{1}{2}, -\frac{1}{2}, \frac{3}{2}, -\frac{1}{2}\rangle_{\lambda=2}$, and $|4\rangle = | \frac{1}{2}, \frac{1}{2}, \frac{3}{2}, -\frac{3}{2}\rangle_{\lambda=2}$, expressed in the same basis.
We perform a basis transformation to the $|S, M_S, I,  M_I\rangle_\lambda$ basis (see \textit{e.g.} \cite{chin2010feshbach,2011NJPh...13h3005I}) and solve the Eq. (\ref{eq:matrixform2}) using the log-derivative method \cite{Johnson1973} to calculate the $S$-matrix and extract the complex scattering length $\eta(B)\equiv a(B) - i b(B)$ in the ultracold limit \cite{Mott:1987,1997CPL...280....5B,chin2010feshbach} as a function of the magnetic field for the scattering energy $E=10$ $\mu$K above the lower asymptote ($\lambda=1$).
The scattering length is shown in Fig. \ref{fig3} and compared to the asymptotic bound-state model (ABM) \cite{chin2010feshbach} with mass-polarization couplings. 

For the selected entrance channel, we find four Feshbach resonances caused by the coupling of the entrance channel $|1\rangle$ and closed channel $|3\rangle$. The long-range form of atom-ion potentials, $V(R)\propto -1/R^4$, supports a higher density of bound states than the atom-atom interaction, contributing to a larger number of resonances. The broadest resonance occurs at about $B_0=322$ G and has a width $\Delta B=3$ mG (Fig. \ref{fig3}). 
We can estimate the CX cross section $\sigma^{\mathrm{cx}}_{1\rightarrow3}$ in the ultracold limit from the expression $\sigma^{\mathrm{cx}}_{1\rightarrow 3} = \frac{\pi}{k^{2}}\left( 1 - |S_{1,3}|^2\right) = \frac{4\pi}{k} b$ \cite{1997CPL...280....5B}, where $k = \sqrt{2 \mu (E - \epsilon_1)}$ and $\epsilon_1(B)$ is the threshold energy of the channel $|1\rangle$.
For the conditions considered, the CX cross section at the full-width-half-maximum (FWHM) of the resonance at $B=B_0$ becomes  $\sigma^{\mathrm{cx}}_{1\rightarrow3} \approx 10^{-14}$ cm$^2$ (for $k$ corresponding to 10 $\mu$K), where we assume $b = 0.02$ a.u. over the resonance width $\Delta B=3$ mG.
This value compares to the elastic cross section $\sigma^{\mathrm{el}} = 4\pi \left(a^2 + b^2\right) \approx 10^{-12}$ cm$^2$ for the same conditions. 
Therefore, at the Feshbach resonance, about 1\% of collisions will result in charge exchange. 
The resulting CX rate at ultracold conditions, $K_\mathrm{cx}(B) = 4 \pi b(B) / \mu$ is given in Fig. \ref{fig3} \textit{(bottom panel)} for the lowest-field resonance. At the resonance, the CX rate increases from negligible values to about $7 \times 10^{-12}$ cm$^3$s$^{-1}$. Similar results are obtained for higher $B$-field resonances and for other channels. 
At the given energy, the contributions from higher partial waves (first four partial waves are significant) do not visibly affect the resonance effects.
Note that the position and width of the Feshbach resonance at 322 G discussed above are very sensitive on the details of the potential energy curves and presently impossible to determine accurately from \textit{ab-initio} calculations alone. Consequently, while we know that the resonances will be present, as well as their number, their properties could be different in reality.

The CX discussed above takes place between an open channel ($|1\rangle$) and a closed channel ($|3\rangle$) in different asymptotic charge arrangements (labeled as $\lambda=1$ and $\lambda=2$, and corresponding asymptotically to $^9$Be$^+$+$^{10}$Be and $^9$Be+$^{10}$Be$^+$, respectively). It results in population transfer to a loosely bound state of the molecular ion ($^9$Be$^{10}$Be)$^+$ that is not asymptotically free (see Fig. \ref{fig0}), but rather ``quasi-free" since such states near the dissociation limit of $R^{-4}$ potential are extremely extended (up to thousands of bohr radii). 
Therefore, strictly speaking, the ``CX rate" for these states corresponds to the rate of formation of a magnetically induced Feshbach resonance rather than to the rate of CX between two asymptotically open channels of different charge arrangement. 
The picture changes qualitatively at very high magnetic fields, above 5000 G (Fig. \ref{fig3}), where more than one channel of different charge arrangement is open and the CX takes place between the asymptotes with different charge arrangement separated at infinity by the mass polarization term. We did not study this regime in detail due to very high magnetic fields involved. However, it is conceivable that for different heteroisotopic pairs significantly smaller fields would suffice.

In summary, using $^9$Be$^+$ and $^{10}$Be as a model system, we have analyzed the prospects of controlling charge exchange in ultracold atom-ion collisions via an external magnetic field. We have shown that weak couplings due to mass polarization terms are responsible for the appearance of Feshbach resonances between two lowest electronic states with different asymptotic charge arrangements ($^9$Be$^+$+$^{10}$Be and $^9$Be+$^{10}$Be$^+$), leading to resonantly enhanced charge exchange between the entrance channel and near-threshold bound vibrational states of the ``Feshbach molecular ion''. 
These states are very loosely bound and extend spatially to thousands of Bohr radii due to the long-range character of atom-ion interaction.
The reported Feshbach resonances exhibit the usual properties as in neutral systems, and, for the analyzed system, allow the CX cross sections to be tuned from zero to about 1\% of the elastic cross section value. Even though elastic scattering remains the dominant process at ultracold temperatures, the CX rate coefficient can be tuned to significant values of $10^{-12}$ cm$^3$/s or greater, that are experimentally observable. A possible CX detection scheme could use multi-photon ionization of the resonantly populated bound states. Alternatively, in analogy to magnetoassociation of neutral dimers, performing a time-dependent sweep of the magnetic field across the resonance would likely retain some of the resonantly transferred population in the bound state of the molecular ion.
This previously unexplored mechanism involving Feshbach resonances has possible applications in studies of charge mobility and diffusion in ultracold gases (see \textit{e.g.}, \cite{2000PhRvL..85.5316C}) and physical systems involving one or more ions in a BEC \cite{2002PhRvL..89i3001C,2010Natur.464..388Z,2013Natur.502..664B,2015PhRvL.114x3003W}. 

In concluding, we note that the analyzed system likely exhibits a reduction of the resonant coupling strength due to the fact that the $B^2\Sigma^+$ state is weakly repulsive and, at ultracold conditions, prevents the interaction to take place at smaller nuclear separations where the mass polarization term is stronger. Consequently, Feshbach-resonant charge exchange cross sections in atom-ion systems with similar electronic configurations, such as (CaCa)$^+$, (MgMg)$^+$ or (SrSr)$^+$, are likely to be larger \cite{2012CPL...542..138B,2013MolPh.111.2292L}. 
In a broader sense, if more than one open channel is present, the described CX process involving different isotopes can be understood as a chemical reaction with the energy barrier equal to the isotope shift. In fact, as opposed to previous studies of the effect of Feshbach resonances within a given charge arrangement on radiative charge exchange, this work can be seen as an illustration of a controlled chemical reaction, albeit a trivial one, where it becomes possible to tune the probability for selecting a particular exit channel, or a reaction product, via an external magnetic field.

\begin{acknowledgments}
This work was partially supported by the MURI U.S. Army Research Office grant number W911NF-14-1-0378 (MG), and by the PIF program of the National Science Foundation grant number PHY-1415560 (RC). MG was partially supported by an appointment to the NASA Postdoctoral Program at the NASA Ames Research Center, administered by Universities Space Research Association under contract with NASA.
\end{acknowledgments}

\bibliography{refs_cx}

\begin{thebibliography}{68}
\expandafter\ifx\csname natexlab\endcsname\relax\def\natexlab#1{#1}\fi
\expandafter\ifx\csname bibnamefont\endcsname\relax
  \def\bibnamefont#1{#1}\fi
\expandafter\ifx\csname bibfnamefont\endcsname\relax
  \def\bibfnamefont#1{#1}\fi
\expandafter\ifx\csname citenamefont\endcsname\relax
  \def\citenamefont#1{#1}\fi
\expandafter\ifx\csname url\endcsname\relax
  \def\url#1{\texttt{#1}}\fi
\expandafter\ifx\csname urlprefix\endcsname\relax\def\urlprefix{URL }\fi
\providecommand{\bibinfo}[2]{#2}
\providecommand{\eprint}[2][]{\url{#2}}

\bibitem[{\citenamefont{{C{\^o}t{\'e}} and
  {Dalgarno}}(2000)}]{2000PhRvA..62a2709C}
\bibinfo{author}{\bibfnamefont{R.}~\bibnamefont{{C{\^o}t{\'e}}}}
  \bibnamefont{and}
  \bibinfo{author}{\bibfnamefont{A.}~\bibnamefont{{Dalgarno}}},
  \bibinfo{journal}{\pra} \textbf{\bibinfo{volume}{62}}, \bibinfo{eid}{012709}
  (\bibinfo{year}{2000}).

\bibitem[{\citenamefont{{C{\^o}t{\'e}}}(2000)}]{2000PhRvL..85.5316C}
\bibinfo{author}{\bibfnamefont{R.}~\bibnamefont{{C{\^o}t{\'e}}}},
  \bibinfo{journal}{\prl} \textbf{\bibinfo{volume}{85}}, \bibinfo{pages}{5316}
  (\bibinfo{year}{2000}).

\bibitem[{\citenamefont{{Makarov} et~al.}(2003)\citenamefont{{Makarov},
  {C{\^o}t{\'e}}, {Michels}, and {Smith}}}]{2003PhRvA..67d2705M}
\bibinfo{author}{\bibfnamefont{O.~P.} \bibnamefont{{Makarov}}},
  \bibinfo{author}{\bibfnamefont{R.}~\bibnamefont{{C{\^o}t{\'e}}}},
  \bibinfo{author}{\bibfnamefont{H.}~\bibnamefont{{Michels}}},
  \bibnamefont{and} \bibinfo{author}{\bibfnamefont{W.~W.}
  \bibnamefont{{Smith}}}, \bibinfo{journal}{\pra}
  \textbf{\bibinfo{volume}{67}}, \bibinfo{eid}{042705} (\bibinfo{year}{2003}).

\bibitem[{\citenamefont{{Willitsch}}(2014)}]{2014arXiv1401.1699W}
\bibinfo{author}{\bibfnamefont{S.}~\bibnamefont{{Willitsch}}},
  \bibinfo{journal}{ArXiv e-prints}  (\bibinfo{year}{2014}),
  \eprint{1401.1699}.

\bibitem[{\citenamefont{{H{\"a}rter} and {Hecker
  Denschlag}}(2014)}]{2014ConPh..55...33H}
\bibinfo{author}{\bibfnamefont{A.}~\bibnamefont{{H{\"a}rter}}}
  \bibnamefont{and} \bibinfo{author}{\bibfnamefont{J.}~\bibnamefont{{Hecker
  Denschlag}}}, \bibinfo{journal}{Contemp. Phys.}
  \textbf{\bibinfo{volume}{55}}, \bibinfo{pages}{33} (\bibinfo{year}{2014}).

\bibitem[{\citenamefont{{C{\^o}t{\'e}}}(2016)}]{2016AAMOP..65...67C}
\bibinfo{author}{\bibfnamefont{R.}~\bibnamefont{{C{\^o}t{\'e}}}},
  \bibinfo{journal}{Adv. At. Mol. Opt. Phy.} \textbf{\bibinfo{volume}{65}},
  \bibinfo{pages}{67} (\bibinfo{year}{2016}).

\bibitem[{\citenamefont{{Huber} et~al.}(2014)\citenamefont{{Huber},
  {Lambrecht}, {Schmidt}, {Karpa}, and {Schaetz}}}]{2014NatCo...5E5587H}
\bibinfo{author}{\bibfnamefont{T.}~\bibnamefont{{Huber}}},
  \bibinfo{author}{\bibfnamefont{A.}~\bibnamefont{{Lambrecht}}},
  \bibinfo{author}{\bibfnamefont{J.}~\bibnamefont{{Schmidt}}},
  \bibinfo{author}{\bibfnamefont{L.}~\bibnamefont{{Karpa}}}, \bibnamefont{and}
  \bibinfo{author}{\bibfnamefont{T.}~\bibnamefont{{Schaetz}}},
  \bibinfo{journal}{Nat. Commun.} \textbf{\bibinfo{volume}{5}},
  \bibinfo{eid}{5587} (\bibinfo{year}{2014}).

\bibitem[{\citenamefont{{Eberle} et~al.}(2015)\citenamefont{{Eberle},
  {D{\"o}rfler}, {von Planta}, {Ravi}, {Haas}, {Zhang}, {van de Meerakker}, and
  {Willitsch}}}]{2015JPhCS.635a2012E}
\bibinfo{author}{\bibfnamefont{P.}~\bibnamefont{{Eberle}}},
  \bibinfo{author}{\bibfnamefont{A.~D.} \bibnamefont{{D{\"o}rfler}}},
  \bibinfo{author}{\bibfnamefont{C.}~\bibnamefont{{von Planta}}},
  \bibinfo{author}{\bibfnamefont{K.}~\bibnamefont{{Ravi}}},
  \bibinfo{author}{\bibfnamefont{D.}~\bibnamefont{{Haas}}},
  \bibinfo{author}{\bibfnamefont{D.}~\bibnamefont{{Zhang}}},
  \bibinfo{author}{\bibfnamefont{S.~Y.~T.} \bibnamefont{{van de Meerakker}}},
  \bibnamefont{and}
  \bibinfo{author}{\bibfnamefont{S.}~\bibnamefont{{Willitsch}}},
  \bibinfo{journal}{J. Phys. Conf. Ser.} \textbf{\bibinfo{volume}{635}},
  \bibinfo{eid}{012012} (\bibinfo{year}{2015}).

\bibitem[{\citenamefont{{Schowalter} et~al.}(2016)\citenamefont{{Schowalter},
  {Dunning}, {Chen}, {Puri}, {Schneider}, and {Hudson}}}]{2016NatCo...712448S}
\bibinfo{author}{\bibfnamefont{S.~J.} \bibnamefont{{Schowalter}}},
  \bibinfo{author}{\bibfnamefont{A.~J.} \bibnamefont{{Dunning}}},
  \bibinfo{author}{\bibfnamefont{K.}~\bibnamefont{{Chen}}},
  \bibinfo{author}{\bibfnamefont{P.}~\bibnamefont{{Puri}}},
  \bibinfo{author}{\bibfnamefont{C.}~\bibnamefont{{Schneider}}},
  \bibnamefont{and} \bibinfo{author}{\bibfnamefont{E.~R.}
  \bibnamefont{{Hudson}}}, \bibinfo{journal}{Nat. Commun.}
  \textbf{\bibinfo{volume}{7}}, \bibinfo{eid}{12448} (\bibinfo{year}{2016}).

\bibitem[{\citenamefont{{H{\"o}ltkemeier}
  et~al.}(2016)\citenamefont{{H{\"o}ltkemeier}, {Weckesser},
  {L{\'o}pez-Carrera}, and {Weidem{\"u}ller}}}]{2016PhRvL.116w3003H}
\bibinfo{author}{\bibfnamefont{B.}~\bibnamefont{{H{\"o}ltkemeier}}},
  \bibinfo{author}{\bibfnamefont{P.}~\bibnamefont{{Weckesser}}},
  \bibinfo{author}{\bibfnamefont{H.}~\bibnamefont{{L{\'o}pez-Carrera}}},
  \bibnamefont{and}
  \bibinfo{author}{\bibfnamefont{M.}~\bibnamefont{{Weidem{\"u}ller}}},
  \bibinfo{journal}{\prl} \textbf{\bibinfo{volume}{116}}, \bibinfo{eid}{233003}
  (\bibinfo{year}{2016}).

\bibitem[{\citenamefont{{Rakshit} et~al.}(2016)\citenamefont{{Rakshit},
  {Ghanmi}, {Berriche}, and {Deb}}}]{2016JPhB...49j5202R}
\bibinfo{author}{\bibfnamefont{A.}~\bibnamefont{{Rakshit}}},
  \bibinfo{author}{\bibfnamefont{C.}~\bibnamefont{{Ghanmi}}},
  \bibinfo{author}{\bibfnamefont{H.}~\bibnamefont{{Berriche}}},
  \bibnamefont{and} \bibinfo{author}{\bibfnamefont{B.}~\bibnamefont{{Deb}}},
  \bibinfo{journal}{J. Phys. B} \textbf{\bibinfo{volume}{49}},
  \bibinfo{eid}{105202} (\bibinfo{year}{2016}).

\bibitem[{\citenamefont{{Smith} et~al.}(2005)\citenamefont{{Smith}, {Makarov},
  and {Lin}}}]{2005JMOp...52.2253S}
\bibinfo{author}{\bibfnamefont{W.~W.} \bibnamefont{{Smith}}},
  \bibinfo{author}{\bibfnamefont{O.~P.} \bibnamefont{{Makarov}}},
  \bibnamefont{and} \bibinfo{author}{\bibfnamefont{J.}~\bibnamefont{{Lin}}},
  \bibinfo{journal}{J. Mod. Opt.} \textbf{\bibinfo{volume}{52}},
  \bibinfo{pages}{2253} (\bibinfo{year}{2005}).

\bibitem[{\citenamefont{{Grier} et~al.}(2009)\citenamefont{{Grier}, {Cetina},
  {Oru{\v c}evi{\'c}}, and {Vuleti{\'c}}}}]{2009PhRvL.102v3201G}
\bibinfo{author}{\bibfnamefont{A.~T.} \bibnamefont{{Grier}}},
  \bibinfo{author}{\bibfnamefont{M.}~\bibnamefont{{Cetina}}},
  \bibinfo{author}{\bibfnamefont{F.}~\bibnamefont{{Oru{\v c}evi{\'c}}}},
  \bibnamefont{and}
  \bibinfo{author}{\bibfnamefont{V.}~\bibnamefont{{Vuleti{\'c}}}},
  \bibinfo{journal}{\prl} \textbf{\bibinfo{volume}{102}}, \bibinfo{eid}{223201}
  (\bibinfo{year}{2009}).

\bibitem[{\citenamefont{{Tong} et~al.}(2010)\citenamefont{{Tong}, {Winney}, and
  {Willitsch}}}]{2010PhRvL.105n3001T}
\bibinfo{author}{\bibfnamefont{X.}~\bibnamefont{{Tong}}},
  \bibinfo{author}{\bibfnamefont{A.~H.} \bibnamefont{{Winney}}},
  \bibnamefont{and}
  \bibinfo{author}{\bibfnamefont{S.}~\bibnamefont{{Willitsch}}},
  \bibinfo{journal}{\prl} \textbf{\bibinfo{volume}{105}}, \bibinfo{eid}{143001}
  (\bibinfo{year}{2010}).

\bibitem[{\citenamefont{{Zipkes} et~al.}(2010)\citenamefont{{Zipkes}, {Palzer},
  {Sias}, and {K{\"o}hl}}}]{2010Natur.464..388Z}
\bibinfo{author}{\bibfnamefont{C.}~\bibnamefont{{Zipkes}}},
  \bibinfo{author}{\bibfnamefont{S.}~\bibnamefont{{Palzer}}},
  \bibinfo{author}{\bibfnamefont{C.}~\bibnamefont{{Sias}}}, \bibnamefont{and}
  \bibinfo{author}{\bibfnamefont{M.}~\bibnamefont{{K{\"o}hl}}},
  \bibinfo{journal}{\nat} \textbf{\bibinfo{volume}{464}}, \bibinfo{pages}{388}
  (\bibinfo{year}{2010}).

\bibitem[{\citenamefont{{Rellergert} et~al.}(2011)\citenamefont{{Rellergert},
  {Sullivan}, {Kotochigova}, {Petrov}, {Chen}, {Schowalter}, and
  {Hudson}}}]{2011PhRvL.107x3201R}
\bibinfo{author}{\bibfnamefont{W.~G.} \bibnamefont{{Rellergert}}},
  \bibinfo{author}{\bibfnamefont{S.~T.} \bibnamefont{{Sullivan}}},
  \bibinfo{author}{\bibfnamefont{S.}~\bibnamefont{{Kotochigova}}},
  \bibinfo{author}{\bibfnamefont{A.}~\bibnamefont{{Petrov}}},
  \bibinfo{author}{\bibfnamefont{K.}~\bibnamefont{{Chen}}},
  \bibinfo{author}{\bibfnamefont{S.~J.} \bibnamefont{{Schowalter}}},
  \bibnamefont{and} \bibinfo{author}{\bibfnamefont{E.~R.}
  \bibnamefont{{Hudson}}}, \bibinfo{journal}{\prl}
  \textbf{\bibinfo{volume}{107}}, \bibinfo{eid}{243201} (\bibinfo{year}{2011}).

\bibitem[{\citenamefont{{Sivarajah} et~al.}(2012)\citenamefont{{Sivarajah},
  {Goodman}, {Wells}, {Narducci}, and {Smith}}}]{2012PhRvA..86f3419S}
\bibinfo{author}{\bibfnamefont{I.}~\bibnamefont{{Sivarajah}}},
  \bibinfo{author}{\bibfnamefont{D.~S.} \bibnamefont{{Goodman}}},
  \bibinfo{author}{\bibfnamefont{J.~E.} \bibnamefont{{Wells}}},
  \bibinfo{author}{\bibfnamefont{F.~A.} \bibnamefont{{Narducci}}},
  \bibnamefont{and} \bibinfo{author}{\bibfnamefont{W.~W.}
  \bibnamefont{{Smith}}}, \bibinfo{journal}{\pra}
  \textbf{\bibinfo{volume}{86}}, \bibinfo{eid}{063419} (\bibinfo{year}{2012}).

\bibitem[{\citenamefont{{Ravi} et~al.}(2012)\citenamefont{{Ravi}, {Lee},
  {Sharma}, {Werth}, and {Rangwala}}}]{2012NatCo...3E1126R}
\bibinfo{author}{\bibfnamefont{K.}~\bibnamefont{{Ravi}}},
  \bibinfo{author}{\bibfnamefont{S.}~\bibnamefont{{Lee}}},
  \bibinfo{author}{\bibfnamefont{A.}~\bibnamefont{{Sharma}}},
  \bibinfo{author}{\bibfnamefont{G.}~\bibnamefont{{Werth}}}, \bibnamefont{and}
  \bibinfo{author}{\bibfnamefont{S.~A.} \bibnamefont{{Rangwala}}},
  \bibinfo{journal}{Nat. Commun.} \textbf{\bibinfo{volume}{3}},
  \bibinfo{eid}{1126} (\bibinfo{year}{2012}).

\bibitem[{\citenamefont{{Blythe} et~al.}(2005)\citenamefont{{Blythe}, {Roth},
  {Fr{\"o}hlich}, {Wenz}, and {Schiller}}}]{2005PhRvL..95r3002B}
\bibinfo{author}{\bibfnamefont{P.}~\bibnamefont{{Blythe}}},
  \bibinfo{author}{\bibfnamefont{B.}~\bibnamefont{{Roth}}},
  \bibinfo{author}{\bibfnamefont{U.}~\bibnamefont{{Fr{\"o}hlich}}},
  \bibinfo{author}{\bibfnamefont{H.}~\bibnamefont{{Wenz}}}, \bibnamefont{and}
  \bibinfo{author}{\bibfnamefont{S.}~\bibnamefont{{Schiller}}},
  \bibinfo{journal}{\prl} \textbf{\bibinfo{volume}{95}}, \bibinfo{eid}{183002}
  (\bibinfo{year}{2005}).

\bibitem[{\citenamefont{{Roth} et~al.}(2008)\citenamefont{{Roth}, {Offenberg},
  {Zhang}, and {Schiller}}}]{2008PhRvA..78d2709R}
\bibinfo{author}{\bibfnamefont{B.}~\bibnamefont{{Roth}}},
  \bibinfo{author}{\bibfnamefont{D.}~\bibnamefont{{Offenberg}}},
  \bibinfo{author}{\bibfnamefont{C.~B.} \bibnamefont{{Zhang}}},
  \bibnamefont{and}
  \bibinfo{author}{\bibfnamefont{S.}~\bibnamefont{{Schiller}}},
  \bibinfo{journal}{\pra} \textbf{\bibinfo{volume}{78}}, \bibinfo{eid}{042709}
  (\bibinfo{year}{2008}).

\bibitem[{\citenamefont{{Ratschbacher}
  et~al.}(2012)\citenamefont{{Ratschbacher}, {Zipkes}, {Sias}, and
  {K{\"o}hl}}}]{2012NatPh...8..649R}
\bibinfo{author}{\bibfnamefont{L.}~\bibnamefont{{Ratschbacher}}},
  \bibinfo{author}{\bibfnamefont{C.}~\bibnamefont{{Zipkes}}},
  \bibinfo{author}{\bibfnamefont{C.}~\bibnamefont{{Sias}}}, \bibnamefont{and}
  \bibinfo{author}{\bibfnamefont{M.}~\bibnamefont{{K{\"o}hl}}},
  \bibinfo{journal}{Nat. Phys.} \textbf{\bibinfo{volume}{8}},
  \bibinfo{pages}{649} (\bibinfo{year}{2012}).

\bibitem[{\citenamefont{{Haze} et~al.}(2015)\citenamefont{{Haze}, {Saito},
  {Fujinaga}, and {Mukaiyama}}}]{2015PhRvA..91c2709H}
\bibinfo{author}{\bibfnamefont{S.}~\bibnamefont{{Haze}}},
  \bibinfo{author}{\bibfnamefont{R.}~\bibnamefont{{Saito}}},
  \bibinfo{author}{\bibfnamefont{M.}~\bibnamefont{{Fujinaga}}},
  \bibnamefont{and}
  \bibinfo{author}{\bibfnamefont{T.}~\bibnamefont{{Mukaiyama}}},
  \bibinfo{journal}{\pra} \textbf{\bibinfo{volume}{91}}, \bibinfo{eid}{032709}
  (\bibinfo{year}{2015}).

\bibitem[{\citenamefont{{Kr{\"u}kow} et~al.}(2016)\citenamefont{{Kr{\"u}kow},
  {Mohammadi}, {H{\"a}rter}, and {Hecker Denschlag}}}]{2016PhRvA..94c0701K}
\bibinfo{author}{\bibfnamefont{A.}~\bibnamefont{{Kr{\"u}kow}}},
  \bibinfo{author}{\bibfnamefont{A.}~\bibnamefont{{Mohammadi}}},
  \bibinfo{author}{\bibfnamefont{A.}~\bibnamefont{{H{\"a}rter}}},
  \bibnamefont{and} \bibinfo{author}{\bibfnamefont{J.}~\bibnamefont{{Hecker
  Denschlag}}}, \bibinfo{journal}{\pra} \textbf{\bibinfo{volume}{94}},
  \bibinfo{eid}{030701} (\bibinfo{year}{2016}).

\bibitem[{\citenamefont{{Bloom} et~al.}(2014)\citenamefont{{Bloom},
  {Nicholson}, {Williams}, {Campbell}, {Bishof}, {Zhang}, {Zhang}, {Bromley},
  and {Ye}}}]{2014Natur.506...71B}
\bibinfo{author}{\bibfnamefont{B.~J.} \bibnamefont{{Bloom}}},
  \bibinfo{author}{\bibfnamefont{T.~L.} \bibnamefont{{Nicholson}}},
  \bibinfo{author}{\bibfnamefont{J.~R.} \bibnamefont{{Williams}}},
  \bibinfo{author}{\bibfnamefont{S.~L.} \bibnamefont{{Campbell}}},
  \bibinfo{author}{\bibfnamefont{M.}~\bibnamefont{{Bishof}}},
  \bibinfo{author}{\bibfnamefont{X.}~\bibnamefont{{Zhang}}},
  \bibinfo{author}{\bibfnamefont{W.}~\bibnamefont{{Zhang}}},
  \bibinfo{author}{\bibfnamefont{S.~L.} \bibnamefont{{Bromley}}},
  \bibnamefont{and} \bibinfo{author}{\bibfnamefont{J.}~\bibnamefont{{Ye}}},
  \bibinfo{journal}{Nature} \textbf{\bibinfo{volume}{506}}, \bibinfo{pages}{71}
  (\bibinfo{year}{2014}).

\bibitem[{\citenamefont{{C{\^o}t{\'e}}
  et~al.}(2002)\citenamefont{{C{\^o}t{\'e}}, {Kharchenko}, and
  {Lukin}}}]{2002PhRvL..89i3001C}
\bibinfo{author}{\bibfnamefont{R.}~\bibnamefont{{C{\^o}t{\'e}}}},
  \bibinfo{author}{\bibfnamefont{V.}~\bibnamefont{{Kharchenko}}},
  \bibnamefont{and} \bibinfo{author}{\bibfnamefont{M.~D.}
  \bibnamefont{{Lukin}}}, \bibinfo{journal}{\prl}
  \textbf{\bibinfo{volume}{89}}, \bibinfo{eid}{093001} (\bibinfo{year}{2002}).

\bibitem[{\citenamefont{{Schmid} et~al.}(2010)\citenamefont{{Schmid},
  {H{\"a}rter}, and {Denschlag}}}]{2010PhRvL.105m3202S}
\bibinfo{author}{\bibfnamefont{S.}~\bibnamefont{{Schmid}}},
  \bibinfo{author}{\bibfnamefont{A.}~\bibnamefont{{H{\"a}rter}}},
  \bibnamefont{and} \bibinfo{author}{\bibfnamefont{J.~H.}
  \bibnamefont{{Denschlag}}}, \bibinfo{journal}{Phys. Rev. Lett.}
  \textbf{\bibinfo{volume}{105}}, \bibinfo{eid}{133202} (\bibinfo{year}{2010}).

\bibitem[{\citenamefont{{Gerritsma} et~al.}(2012)\citenamefont{{Gerritsma},
  {Negretti}, {Doerk}, {Idziaszek}, {Calarco}, and
  {Schmidt-Kaler}}}]{2012PhRvL.109h0402G}
\bibinfo{author}{\bibfnamefont{R.}~\bibnamefont{{Gerritsma}}},
  \bibinfo{author}{\bibfnamefont{A.}~\bibnamefont{{Negretti}}},
  \bibinfo{author}{\bibfnamefont{H.}~\bibnamefont{{Doerk}}},
  \bibinfo{author}{\bibfnamefont{Z.}~\bibnamefont{{Idziaszek}}},
  \bibinfo{author}{\bibfnamefont{T.}~\bibnamefont{{Calarco}}},
  \bibnamefont{and}
  \bibinfo{author}{\bibfnamefont{F.}~\bibnamefont{{Schmidt-Kaler}}},
  \bibinfo{journal}{\prl} \textbf{\bibinfo{volume}{109}}, \bibinfo{eid}{080402}
  (\bibinfo{year}{2012}).

\bibitem[{\citenamefont{{Bissbort} et~al.}(2013)\citenamefont{{Bissbort},
  {Cocks}, {Negretti}, {Idziaszek}, {Calarco}, {Schmidt-Kaler}, {Hofstetter},
  and {Gerritsma}}}]{2013PhRvL.111h0501B}
\bibinfo{author}{\bibfnamefont{U.}~\bibnamefont{{Bissbort}}},
  \bibinfo{author}{\bibfnamefont{D.}~\bibnamefont{{Cocks}}},
  \bibinfo{author}{\bibfnamefont{A.}~\bibnamefont{{Negretti}}},
  \bibinfo{author}{\bibfnamefont{Z.}~\bibnamefont{{Idziaszek}}},
  \bibinfo{author}{\bibfnamefont{T.}~\bibnamefont{{Calarco}}},
  \bibinfo{author}{\bibfnamefont{F.}~\bibnamefont{{Schmidt-Kaler}}},
  \bibinfo{author}{\bibfnamefont{W.}~\bibnamefont{{Hofstetter}}},
  \bibnamefont{and}
  \bibinfo{author}{\bibfnamefont{R.}~\bibnamefont{{Gerritsma}}},
  \bibinfo{journal}{\prl} \textbf{\bibinfo{volume}{111}}, \bibinfo{eid}{080501}
  (\bibinfo{year}{2013}).

\bibitem[{\citenamefont{{Schurer} et~al.}(2014)\citenamefont{{Schurer},
  {Schmelcher}, and {Negretti}}}]{2014PhRvA..90c3601S}
\bibinfo{author}{\bibfnamefont{J.~M.} \bibnamefont{{Schurer}}},
  \bibinfo{author}{\bibfnamefont{P.}~\bibnamefont{{Schmelcher}}},
  \bibnamefont{and}
  \bibinfo{author}{\bibfnamefont{A.}~\bibnamefont{{Negretti}}},
  \bibinfo{journal}{\pra} \textbf{\bibinfo{volume}{90}}, \bibinfo{eid}{033601}
  (\bibinfo{year}{2014}).

\bibitem[{\citenamefont{{Schurer} et~al.}(2016)\citenamefont{{Schurer},
  {Gerritsma}, {Schmelcher}, and {Negretti}}}]{2016PhRvA..93f3602S}
\bibinfo{author}{\bibfnamefont{J.~M.} \bibnamefont{{Schurer}}},
  \bibinfo{author}{\bibfnamefont{R.}~\bibnamefont{{Gerritsma}}},
  \bibinfo{author}{\bibfnamefont{P.}~\bibnamefont{{Schmelcher}}},
  \bibnamefont{and}
  \bibinfo{author}{\bibfnamefont{A.}~\bibnamefont{{Negretti}}},
  \bibinfo{journal}{\pra} \textbf{\bibinfo{volume}{93}}, \bibinfo{eid}{063602}
  (\bibinfo{year}{2016}).

\bibitem[{\citenamefont{{Secker} et~al.}(2016)\citenamefont{{Secker},
  {Gerritsma}, {Glaetzle}, and {Negretti}}}]{2016PhRvA..94a3420S}
\bibinfo{author}{\bibfnamefont{T.}~\bibnamefont{{Secker}}},
  \bibinfo{author}{\bibfnamefont{R.}~\bibnamefont{{Gerritsma}}},
  \bibinfo{author}{\bibfnamefont{A.~W.} \bibnamefont{{Glaetzle}}},
  \bibnamefont{and}
  \bibinfo{author}{\bibfnamefont{A.}~\bibnamefont{{Negretti}}},
  \bibinfo{journal}{\pra} \textbf{\bibinfo{volume}{94}}, \bibinfo{eid}{013420}
  (\bibinfo{year}{2016}).

\bibitem[{\citenamefont{{Doerk} et~al.}(2010)\citenamefont{{Doerk},
  {Idziaszek}, and {Calarco}}}]{2010PhRvA..81a2708D}
\bibinfo{author}{\bibfnamefont{H.}~\bibnamefont{{Doerk}}},
  \bibinfo{author}{\bibfnamefont{Z.}~\bibnamefont{{Idziaszek}}},
  \bibnamefont{and}
  \bibinfo{author}{\bibfnamefont{T.}~\bibnamefont{{Calarco}}},
  \bibinfo{journal}{\pra} \textbf{\bibinfo{volume}{81}}, \bibinfo{eid}{012708}
  (\bibinfo{year}{2010}).

\bibitem[{\citenamefont{{Wineland} and
  {Leibfried}}(2011)}]{2011LaPhL...8..188W}
\bibinfo{author}{\bibfnamefont{D.~J.} \bibnamefont{{Wineland}}}
  \bibnamefont{and}
  \bibinfo{author}{\bibfnamefont{D.}~\bibnamefont{{Leibfried}}},
  \bibinfo{journal}{Laser Phys. Lett.} \textbf{\bibinfo{volume}{8}},
  \bibinfo{eid}{188} (\bibinfo{year}{2011}).

\bibitem[{\citenamefont{{Sayfutyarova}
  et~al.}(2013)\citenamefont{{Sayfutyarova}, {Buchachenko}, {Yakovleva}, and
  {Belyaev}}}]{2013PhRvA..87e2717S}
\bibinfo{author}{\bibfnamefont{E.~R.} \bibnamefont{{Sayfutyarova}}},
  \bibinfo{author}{\bibfnamefont{A.~A.} \bibnamefont{{Buchachenko}}},
  \bibinfo{author}{\bibfnamefont{S.~A.} \bibnamefont{{Yakovleva}}},
  \bibnamefont{and} \bibinfo{author}{\bibfnamefont{A.~K.}
  \bibnamefont{{Belyaev}}}, \bibinfo{journal}{\pra}
  \textbf{\bibinfo{volume}{87}}, \bibinfo{eid}{052717} (\bibinfo{year}{2013}).

\bibitem[{\citenamefont{{McLaughlin} et~al.}(2014)\citenamefont{{McLaughlin},
  {Lamb}, {Lane}, and {McCann}}}]{2014JPhB...47n5201M}
\bibinfo{author}{\bibfnamefont{B.~M.} \bibnamefont{{McLaughlin}}},
  \bibinfo{author}{\bibfnamefont{H.~D.~L.} \bibnamefont{{Lamb}}},
  \bibinfo{author}{\bibfnamefont{I.~C.} \bibnamefont{{Lane}}},
  \bibnamefont{and} \bibinfo{author}{\bibfnamefont{J.~F.}
  \bibnamefont{{McCann}}}, \bibinfo{journal}{J. Phys. B}
  \textbf{\bibinfo{volume}{47}}, \bibinfo{eid}{145201} (\bibinfo{year}{2014}).

\bibitem[{\citenamefont{{Yakovleva} et~al.}(2014)\citenamefont{{Yakovleva},
  {Belyaev}, and {Buchachenko}}}]{2014JPhCS.572a2009Y}
\bibinfo{author}{\bibfnamefont{S.~A.} \bibnamefont{{Yakovleva}}},
  \bibinfo{author}{\bibfnamefont{A.~K.} \bibnamefont{{Belyaev}}},
  \bibnamefont{and} \bibinfo{author}{\bibfnamefont{A.~A.}
  \bibnamefont{{Buchachenko}}}, \bibinfo{journal}{J. Phys. Conf. Ser.}
  \textbf{\bibinfo{volume}{572}}, \bibinfo{eid}{012009} (\bibinfo{year}{2014}).

\bibitem[{\citenamefont{{Tomza} et~al.}(2015)\citenamefont{{Tomza}, {Koch}, and
  {Moszynski}}}]{2015PhRvA..91d2706T}
\bibinfo{author}{\bibfnamefont{M.}~\bibnamefont{{Tomza}}},
  \bibinfo{author}{\bibfnamefont{C.~P.} \bibnamefont{{Koch}}},
  \bibnamefont{and}
  \bibinfo{author}{\bibfnamefont{R.}~\bibnamefont{{Moszynski}}},
  \bibinfo{journal}{\pra} \textbf{\bibinfo{volume}{91}}, \bibinfo{eid}{042706}
  (\bibinfo{year}{2015}).

\bibitem[{\citenamefont{{da Silva} et~al.}(2015)\citenamefont{{da Silva},
  {Raoult}, {Aymar}, and {Dulieu}}}]{2015NJPh...17d5015D}
\bibinfo{author}{\bibfnamefont{H.}~\bibnamefont{{da Silva}},
  \bibfnamefont{Jr.}},
  \bibinfo{author}{\bibfnamefont{M.}~\bibnamefont{{Raoult}}},
  \bibinfo{author}{\bibfnamefont{M.}~\bibnamefont{{Aymar}}}, \bibnamefont{and}
  \bibinfo{author}{\bibfnamefont{O.}~\bibnamefont{{Dulieu}}},
  \bibinfo{journal}{New J. Phys.} \textbf{\bibinfo{volume}{17}},
  \bibinfo{eid}{045015} (\bibinfo{year}{2015}).

\bibitem[{\citenamefont{{Gacesa} et~al.}(2016)\citenamefont{{Gacesa},
  {Montgomery}, {Michels}, and {C{\^o}t{\'e}}}}]{2016PhRvA..94a3407G}
\bibinfo{author}{\bibfnamefont{M.}~\bibnamefont{{Gacesa}}},
  \bibinfo{author}{\bibfnamefont{J.~A.} \bibnamefont{{Montgomery}}},
  \bibinfo{author}{\bibfnamefont{H.~H.} \bibnamefont{{Michels}}},
  \bibnamefont{and}
  \bibinfo{author}{\bibfnamefont{R.}~\bibnamefont{{C{\^o}t{\'e}}}},
  \bibinfo{journal}{\pra} \textbf{\bibinfo{volume}{94}}, \bibinfo{eid}{013407}
  (\bibinfo{year}{2016}).

\bibitem[{\citenamefont{{Saito} et~al.}(2016)\citenamefont{{Saito}, {Haze},
  {Sasakawa}, {Nakai}, {Raoult}, {Da Silva}, {Dulieu}, and
  {Mukaiyama}}}]{2016arXiv160807043S}
\bibinfo{author}{\bibfnamefont{R.}~\bibnamefont{{Saito}}},
  \bibinfo{author}{\bibfnamefont{S.}~\bibnamefont{{Haze}}},
  \bibinfo{author}{\bibfnamefont{M.}~\bibnamefont{{Sasakawa}}},
  \bibinfo{author}{\bibfnamefont{R.}~\bibnamefont{{Nakai}}},
  \bibinfo{author}{\bibfnamefont{M.}~\bibnamefont{{Raoult}}},
  \bibinfo{author}{\bibfnamefont{H.}~\bibnamefont{{Da Silva}},
  \bibfnamefont{Jr.}},
  \bibinfo{author}{\bibfnamefont{O.}~\bibnamefont{{Dulieu}}}, \bibnamefont{and}
  \bibinfo{author}{\bibfnamefont{T.}~\bibnamefont{{Mukaiyama}}},
  \bibinfo{journal}{ArXiv e-prints}  (\bibinfo{year}{2016}),
  \eprint{1608.07043}.

\bibitem[{\citenamefont{{Sardar} et~al.}(2016)\citenamefont{{Sardar}, {Naskar},
  {Pal}, {Berriche}, and {Deb}}}]{2016JPhB...49x5202S}
\bibinfo{author}{\bibfnamefont{D.}~\bibnamefont{{Sardar}}},
  \bibinfo{author}{\bibfnamefont{S.}~\bibnamefont{{Naskar}}},
  \bibinfo{author}{\bibfnamefont{A.}~\bibnamefont{{Pal}}},
  \bibinfo{author}{\bibfnamefont{H.}~\bibnamefont{{Berriche}}},
  \bibnamefont{and} \bibinfo{author}{\bibfnamefont{B.}~\bibnamefont{{Deb}}},
  \bibinfo{journal}{J. Phys. B} \textbf{\bibinfo{volume}{49}},
  \bibinfo{eid}{245202} (\bibinfo{year}{2016}).

\bibitem[{\citenamefont{{Pavlovi\'{c}}
  et~al.}(2005)\citenamefont{{Pavlovi\'{c}}, Krems, C\^ot\'e, and
  Sadeghpour}}]{RbCr2005}
\bibinfo{author}{\bibfnamefont{Z.}~\bibnamefont{{Pavlovi\'{c}}}},
  \bibinfo{author}{\bibfnamefont{R.~V.} \bibnamefont{Krems}},
  \bibinfo{author}{\bibfnamefont{R.}~\bibnamefont{C\^ot\'e}}, \bibnamefont{and}
  \bibinfo{author}{\bibfnamefont{H.~R.} \bibnamefont{Sadeghpour}},
  \bibinfo{journal}{Phys. Rev. A} \textbf{\bibinfo{volume}{71}},
  \bibinfo{pages}{061402} (\bibinfo{year}{2005}).

\bibitem[{\citenamefont{Gacesa et~al.}(2008)\citenamefont{Gacesa, Pellegrini,
  and C\^ot\'e}}]{LiNa2008}
\bibinfo{author}{\bibfnamefont{M.}~\bibnamefont{Gacesa}},
  \bibinfo{author}{\bibfnamefont{P.}~\bibnamefont{Pellegrini}},
  \bibnamefont{and} \bibinfo{author}{\bibfnamefont{R.}~\bibnamefont{C\^ot\'e}},
  \bibinfo{journal}{Phys. Rev. A} \textbf{\bibinfo{volume}{78}},
  \bibinfo{pages}{010701} (\bibinfo{year}{2008}).

\bibitem[{\citenamefont{{Deiglmayr} et~al.}(2009)\citenamefont{{Deiglmayr},
  {Pellegrini}, {Grochola}, {Repp}, {C\^ot\'e}, {Dulieu}, {Wester}, and
  {Weidem\"uller}}}]{LiCs2009}
\bibinfo{author}{\bibfnamefont{J.}~\bibnamefont{{Deiglmayr}}},
  \bibinfo{author}{\bibfnamefont{P.}~\bibnamefont{{Pellegrini}}},
  \bibinfo{author}{\bibfnamefont{A.}~\bibnamefont{{Grochola}}},
  \bibinfo{author}{\bibfnamefont{A.}~\bibnamefont{{Repp}}},
  \bibinfo{author}{\bibfnamefont{R.}~\bibnamefont{{C\^ot\'e}}},
  \bibinfo{author}{\bibfnamefont{O.}~\bibnamefont{{Dulieu}}},
  \bibinfo{author}{\bibfnamefont{R.}~\bibnamefont{{Wester}}}, \bibnamefont{and}
  \bibinfo{author}{\bibfnamefont{M.}~\bibnamefont{{Weidem\"uller}}},
  \bibinfo{journal}{New J. Phys.} \textbf{\bibinfo{volume}{11}},
  \bibinfo{pages}{055034} (\bibinfo{year}{2009}).

\bibitem[{\citenamefont{{Stwalley}}(1976)}]{1976PhRvL..37.1628S}
\bibinfo{author}{\bibfnamefont{W.~C.} \bibnamefont{{Stwalley}}},
  \bibinfo{journal}{\prl} \textbf{\bibinfo{volume}{37}}, \bibinfo{pages}{1628}
  (\bibinfo{year}{1976}).

\bibitem[{\citenamefont{K\"ohler et~al.}(2006)\citenamefont{K\"ohler, G\'oral,
  and Julienne}}]{kohler:1311}
\bibinfo{author}{\bibfnamefont{T.}~\bibnamefont{K\"ohler}},
  \bibinfo{author}{\bibfnamefont{K.}~\bibnamefont{G\'oral}}, \bibnamefont{and}
  \bibinfo{author}{\bibfnamefont{P.~S.} \bibnamefont{Julienne}},
  \bibinfo{journal}{Rev. Mod. Phys.} \textbf{\bibinfo{volume}{78}},
  \bibinfo{eid}{1311} (\bibinfo{year}{2006}).

\bibitem[{\citenamefont{Chin et~al.}(2010)\citenamefont{Chin, Grimm, Julienne,
  and Tiesinga}}]{chin2010feshbach}
\bibinfo{author}{\bibfnamefont{C.}~\bibnamefont{Chin}},
  \bibinfo{author}{\bibfnamefont{R.}~\bibnamefont{Grimm}},
  \bibinfo{author}{\bibfnamefont{P.}~\bibnamefont{Julienne}}, \bibnamefont{and}
  \bibinfo{author}{\bibfnamefont{E.}~\bibnamefont{Tiesinga}},
  \bibinfo{journal}{Rev. Mod. Phys.} \textbf{\bibinfo{volume}{82}},
  \bibinfo{pages}{1225} (\bibinfo{year}{2010}).

\bibitem[{\citenamefont{{Esry} et~al.}(2000)\citenamefont{{Esry}, {Sadeghpour},
  {Wells}, and {Ben-Itzhak}}}]{2000JPhB...33.5329E}
\bibinfo{author}{\bibfnamefont{B.~D.} \bibnamefont{{Esry}}},
  \bibinfo{author}{\bibfnamefont{H.~R.} \bibnamefont{{Sadeghpour}}},
  \bibinfo{author}{\bibfnamefont{E.}~\bibnamefont{{Wells}}}, \bibnamefont{and}
  \bibinfo{author}{\bibfnamefont{I.}~\bibnamefont{{Ben-Itzhak}}},
  \bibinfo{journal}{J. Phys. B} \textbf{\bibinfo{volume}{33}},
  \bibinfo{pages}{5329} (\bibinfo{year}{2000}).

\bibitem[{\citenamefont{{Idziaszek} et~al.}(2011)\citenamefont{{Idziaszek},
  {Simoni}, {Calarco}, and {Julienne}}}]{2011NJPh...13h3005I}
\bibinfo{author}{\bibfnamefont{Z.}~\bibnamefont{{Idziaszek}}},
  \bibinfo{author}{\bibfnamefont{A.}~\bibnamefont{{Simoni}}},
  \bibinfo{author}{\bibfnamefont{T.}~\bibnamefont{{Calarco}}},
  \bibnamefont{and} \bibinfo{author}{\bibfnamefont{P.~S.}
  \bibnamefont{{Julienne}}}, \bibinfo{journal}{New J. Phys.}
  \textbf{\bibinfo{volume}{13}}, \bibinfo{eid}{083005} (\bibinfo{year}{2011}).

\bibitem[{\citenamefont{{Li} et~al.}(2014)\citenamefont{{Li}, {You}, and
  {Gao}}}]{2014PhRvA..89e2704L}
\bibinfo{author}{\bibfnamefont{M.}~\bibnamefont{{Li}}},
  \bibinfo{author}{\bibfnamefont{L.}~\bibnamefont{{You}}}, \bibnamefont{and}
  \bibinfo{author}{\bibfnamefont{B.}~\bibnamefont{{Gao}}},
  \bibinfo{journal}{\pra} \textbf{\bibinfo{volume}{89}}, \bibinfo{eid}{052704}
  (\bibinfo{year}{2014}).

\bibitem[{\citenamefont{{Tomza}}(2015)}]{2015PhRvA..92f2701T}
\bibinfo{author}{\bibfnamefont{M.}~\bibnamefont{{Tomza}}},
  \bibinfo{journal}{\pra} \textbf{\bibinfo{volume}{92}}, \bibinfo{eid}{062701}
  (\bibinfo{year}{2015}).

\bibitem[{\citenamefont{{Sadeghpour} et~al.}(2000)\citenamefont{{Sadeghpour},
  {Bohn}, {Cavagnero}, {Esry}, {Fabrikant}, {Macek}, and
  {Rau}}}]{2000JPhB...33R..93S}
\bibinfo{author}{\bibfnamefont{H.~R.} \bibnamefont{{Sadeghpour}}},
  \bibinfo{author}{\bibfnamefont{J.~L.} \bibnamefont{{Bohn}}},
  \bibinfo{author}{\bibfnamefont{M.~J.} \bibnamefont{{Cavagnero}}},
  \bibinfo{author}{\bibfnamefont{B.~D.} \bibnamefont{{Esry}}},
  \bibinfo{author}{\bibfnamefont{I.~I.} \bibnamefont{{Fabrikant}}},
  \bibinfo{author}{\bibfnamefont{J.~H.} \bibnamefont{{Macek}}},
  \bibnamefont{and} \bibinfo{author}{\bibfnamefont{A.~R.~P.}
  \bibnamefont{{Rau}}}, \bibinfo{journal}{J. Phys. B}
  \textbf{\bibinfo{volume}{33}}, \bibinfo{pages}{R93} (\bibinfo{year}{2000}).

\bibitem[{\citenamefont{{Zhang}
  et~al.}(2009{\natexlab{a}})\citenamefont{{Zhang}, {Dalgarno}, and
  {C{\^o}t{\'e}}}}]{2009PhRvA..80c0703Z}
\bibinfo{author}{\bibfnamefont{P.}~\bibnamefont{{Zhang}}},
  \bibinfo{author}{\bibfnamefont{A.}~\bibnamefont{{Dalgarno}}},
  \bibnamefont{and}
  \bibinfo{author}{\bibfnamefont{R.}~\bibnamefont{{C{\^o}t{\'e}}}},
  \bibinfo{journal}{\pra} \textbf{\bibinfo{volume}{80}}, \bibinfo{eid}{030703}
  (\bibinfo{year}{2009}{\natexlab{a}}).

\bibitem[{\citenamefont{{Bodo} et~al.}(2008)\citenamefont{{Bodo}, {Zhang}, and
  {Dalgarno}}}]{2008NJPh...10c3024B}
\bibinfo{author}{\bibfnamefont{E.}~\bibnamefont{{Bodo}}},
  \bibinfo{author}{\bibfnamefont{P.}~\bibnamefont{{Zhang}}}, \bibnamefont{and}
  \bibinfo{author}{\bibfnamefont{A.}~\bibnamefont{{Dalgarno}}},
  \bibinfo{journal}{New J. Phys.} \textbf{\bibinfo{volume}{10}},
  \bibinfo{eid}{033024} (\bibinfo{year}{2008}).

\bibitem[{\citenamefont{{Zhang}
  et~al.}(2009{\natexlab{b}})\citenamefont{{Zhang}, {Bodo}, and
  {Dalgarno}}}]{2009JPCA..11315085Z}
\bibinfo{author}{\bibfnamefont{P.}~\bibnamefont{{Zhang}}},
  \bibinfo{author}{\bibfnamefont{E.}~\bibnamefont{{Bodo}}}, \bibnamefont{and}
  \bibinfo{author}{\bibfnamefont{A.}~\bibnamefont{{Dalgarno}}},
  \bibinfo{journal}{J. Phys. Chem. A} \textbf{\bibinfo{volume}{113}},
  \bibinfo{pages}{15085} (\bibinfo{year}{2009}{\natexlab{b}}).

\bibitem[{\citenamefont{{Zhang} et~al.}(2011)\citenamefont{{Zhang}, {Dalgarno},
  {C{\^o}t{\'e}}, and {Bodo}}}]{2011PCCP...1319026Z}
\bibinfo{author}{\bibfnamefont{P.}~\bibnamefont{{Zhang}}},
  \bibinfo{author}{\bibfnamefont{A.}~\bibnamefont{{Dalgarno}}},
  \bibinfo{author}{\bibfnamefont{R.}~\bibnamefont{{C{\^o}t{\'e}}}},
  \bibnamefont{and} \bibinfo{author}{\bibfnamefont{E.}~\bibnamefont{{Bodo}}},
  \bibinfo{journal}{Phys. Chem. Chem. Phys.} \textbf{\bibinfo{volume}{13}},
  \bibinfo{pages}{19026} (\bibinfo{year}{2011}).

\bibitem[{\citenamefont{Tscherbul et~al.}(2016)\citenamefont{Tscherbul, Brumer,
  and Buchachenko}}]{PhysRevLett.117.143201}
\bibinfo{author}{\bibfnamefont{T.~V.} \bibnamefont{Tscherbul}},
  \bibinfo{author}{\bibfnamefont{P.}~\bibnamefont{Brumer}}, \bibnamefont{and}
  \bibinfo{author}{\bibfnamefont{A.~A.} \bibnamefont{Buchachenko}},
  \bibinfo{journal}{Phys. Rev. Lett.} \textbf{\bibinfo{volume}{117}},
  \bibinfo{pages}{143201} (\bibinfo{year}{2016}).

\bibitem[{\citenamefont{{Karpiuk} et~al.}(2015)\citenamefont{{Karpiuk},
  {Brewczyk}, {Rz{\c a}{\.z}ewski}, {Gaj}, {Balewski}, {Krupp},
  {Schlagm{\"u}ller}, {L{\"o}w}, {Hofferberth}, and
  {Pfau}}}]{2015NJPh...17e3046K}
\bibinfo{author}{\bibfnamefont{T.}~\bibnamefont{{Karpiuk}}},
  \bibinfo{author}{\bibfnamefont{M.}~\bibnamefont{{Brewczyk}}},
  \bibinfo{author}{\bibfnamefont{K.}~\bibnamefont{{Rz{\c a}{\.z}ewski}}},
  \bibinfo{author}{\bibfnamefont{A.}~\bibnamefont{{Gaj}}},
  \bibinfo{author}{\bibfnamefont{J.~B.} \bibnamefont{{Balewski}}},
  \bibinfo{author}{\bibfnamefont{A.~T.} \bibnamefont{{Krupp}}},
  \bibinfo{author}{\bibfnamefont{M.}~\bibnamefont{{Schlagm{\"u}ller}}},
  \bibinfo{author}{\bibfnamefont{R.}~\bibnamefont{{L{\"o}w}}},
  \bibinfo{author}{\bibfnamefont{S.}~\bibnamefont{{Hofferberth}}},
  \bibnamefont{and} \bibinfo{author}{\bibfnamefont{T.}~\bibnamefont{{Pfau}}},
  \bibinfo{journal}{New J. Phys.} \textbf{\bibinfo{volume}{17}},
  \bibinfo{eid}{053046} (\bibinfo{year}{2015}).

\bibitem[{\citenamefont{{Wang} et~al.}(2015)\citenamefont{{Wang}, {Gacesa}, and
  {C{\^o}t{\'e}}}}]{2015PhRvL.114x3003W}
\bibinfo{author}{\bibfnamefont{J.}~\bibnamefont{{Wang}}},
  \bibinfo{author}{\bibfnamefont{M.}~\bibnamefont{{Gacesa}}}, \bibnamefont{and}
  \bibinfo{author}{\bibfnamefont{R.}~\bibnamefont{{C{\^o}t{\'e}}}},
  \bibinfo{journal}{\prl} \textbf{\bibinfo{volume}{114}}, \bibinfo{eid}{243003}
  (\bibinfo{year}{2015}).

\bibitem[{\citenamefont{Moerdijk et~al.}(1995)\citenamefont{Moerdijk, Verhaar,
  and Axelsson}}]{PhysRevA.51.4852}
\bibinfo{author}{\bibfnamefont{A.~J.} \bibnamefont{Moerdijk}},
  \bibinfo{author}{\bibfnamefont{B.~J.} \bibnamefont{Verhaar}},
  \bibnamefont{and} \bibinfo{author}{\bibfnamefont{A.}~\bibnamefont{Axelsson}},
  \bibinfo{journal}{Phys. Rev. A} \textbf{\bibinfo{volume}{51}},
  \bibinfo{pages}{4852} (\bibinfo{year}{1995}).

\bibitem[{\citenamefont{Grosser et~al.}(1999)\citenamefont{Grosser, Menzel, and
  Belyaev}}]{PhysRevA.59.1309}
\bibinfo{author}{\bibfnamefont{J.}~\bibnamefont{Grosser}},
  \bibinfo{author}{\bibfnamefont{T.}~\bibnamefont{Menzel}}, \bibnamefont{and}
  \bibinfo{author}{\bibfnamefont{A.~K.} \bibnamefont{Belyaev}},
  \bibinfo{journal}{Phys. Rev. A} \textbf{\bibinfo{volume}{59}},
  \bibinfo{pages}{1309} (\bibinfo{year}{1999}).

\bibitem[{\citenamefont{Banerjee et~al.}(2010)\citenamefont{Banerjee, Byrd,
  C{\^o}t{\'e}, Michels, and Montgomery}}]{Banerjee2010208}
\bibinfo{author}{\bibfnamefont{S.}~\bibnamefont{Banerjee}},
  \bibinfo{author}{\bibfnamefont{J.~N.} \bibnamefont{Byrd}},
  \bibinfo{author}{\bibfnamefont{R.}~\bibnamefont{C{\^o}t{\'e}}},
  \bibinfo{author}{\bibfnamefont{H.~H.} \bibnamefont{Michels}},
  \bibnamefont{and} \bibinfo{author}{\bibfnamefont{J.~A.}
  \bibnamefont{Montgomery}}, \bibinfo{journal}{Chem. Phys. Lett.}
  \textbf{\bibinfo{volume}{496}}, \bibinfo{pages}{208 } (\bibinfo{year}{2010}).

\bibitem[{\citenamefont{{Li} et~al.}(2013)\citenamefont{{Li}, {Feng}, {Sun},
  {Zhang}, {Fan}, {Peterson}, {Xie}, and {Schaefer}}}]{2013MolPh.111.2292L}
\bibinfo{author}{\bibfnamefont{H.}~\bibnamefont{{Li}}},
  \bibinfo{author}{\bibfnamefont{H.}~\bibnamefont{{Feng}}},
  \bibinfo{author}{\bibfnamefont{W.}~\bibnamefont{{Sun}}},
  \bibinfo{author}{\bibfnamefont{Y.}~\bibnamefont{{Zhang}}},
  \bibinfo{author}{\bibfnamefont{Q.}~\bibnamefont{{Fan}}},
  \bibinfo{author}{\bibfnamefont{K.~A.} \bibnamefont{{Peterson}}},
  \bibinfo{author}{\bibfnamefont{Y.}~\bibnamefont{{Xie}}}, \bibnamefont{and}
  \bibinfo{author}{\bibfnamefont{H.~F.} \bibnamefont{{Schaefer}},
  \bibfnamefont{III}}, \bibinfo{journal}{Mol. Phys.}
  \textbf{\bibinfo{volume}{111}}, \bibinfo{pages}{2292} (\bibinfo{year}{2013}).

\bibitem[{\citenamefont{Johnson}(1973)}]{Johnson1973}
\bibinfo{author}{\bibfnamefont{B.~R.} \bibnamefont{Johnson}},
  \bibinfo{journal}{J. Comput. Phys.} \textbf{\bibinfo{volume}{13}},
  \bibinfo{pages}{445} (\bibinfo{year}{1973}).

\bibitem[{\citenamefont{Mott and Massey}(1987)}]{Mott:1987}
\bibinfo{author}{\bibfnamefont{N.~F.} \bibnamefont{Mott}} \bibnamefont{and}
  \bibinfo{author}{\bibfnamefont{H.~S.~W.} \bibnamefont{Massey}},
  \emph{\bibinfo{title}{The Theory of Atomic Collisions}}
  (\bibinfo{publisher}{Oxford University Press}, \bibinfo{address}{New York,
  USA}, \bibinfo{year}{1987}), \bibinfo{edition}{3rd} ed.

\bibitem[{\citenamefont{{Balakrishnan}
  et~al.}(1997)\citenamefont{{Balakrishnan}, {Kharchenko}, {Forrey}, and
  {Dalgarno}}}]{1997CPL...280....5B}
\bibinfo{author}{\bibfnamefont{N.}~\bibnamefont{{Balakrishnan}}},
  \bibinfo{author}{\bibfnamefont{V.}~\bibnamefont{{Kharchenko}}},
  \bibinfo{author}{\bibfnamefont{R.~C.} \bibnamefont{{Forrey}}},
  \bibnamefont{and}
  \bibinfo{author}{\bibfnamefont{A.}~\bibnamefont{{Dalgarno}}},
  \bibinfo{journal}{Chem. Phys. Lett.} \textbf{\bibinfo{volume}{280}},
  \bibinfo{pages}{5} (\bibinfo{year}{1997}).

\bibitem[{\citenamefont{{Balewski} et~al.}(2013)\citenamefont{{Balewski},
  {Krupp}, {Gaj}, {Peter}, {B{\"u}chler}, {L{\"o}w}, {Hofferberth}, and
  {Pfau}}}]{2013Natur.502..664B}
\bibinfo{author}{\bibfnamefont{J.~B.} \bibnamefont{{Balewski}}},
  \bibinfo{author}{\bibfnamefont{A.~T.} \bibnamefont{{Krupp}}},
  \bibinfo{author}{\bibfnamefont{A.}~\bibnamefont{{Gaj}}},
  \bibinfo{author}{\bibfnamefont{D.}~\bibnamefont{{Peter}}},
  \bibinfo{author}{\bibfnamefont{H.~P.} \bibnamefont{{B{\"u}chler}}},
  \bibinfo{author}{\bibfnamefont{R.}~\bibnamefont{{L{\"o}w}}},
  \bibinfo{author}{\bibfnamefont{S.}~\bibnamefont{{Hofferberth}}},
  \bibnamefont{and} \bibinfo{author}{\bibfnamefont{T.}~\bibnamefont{{Pfau}}},
  \bibinfo{journal}{Nature} \textbf{\bibinfo{volume}{502}},
  \bibinfo{pages}{664} (\bibinfo{year}{2013}).

\bibitem[{\citenamefont{{Banerjee} et~al.}(2012)\citenamefont{{Banerjee},
  {Montgomery}, {Byrd}, {Michels}, and {C{\^o}t{\'e}}}}]{2012CPL...542..138B}
\bibinfo{author}{\bibfnamefont{S.}~\bibnamefont{{Banerjee}}},
  \bibinfo{author}{\bibfnamefont{J.~A.} \bibnamefont{{Montgomery}}},
  \bibinfo{author}{\bibfnamefont{J.~N.} \bibnamefont{{Byrd}}},
  \bibinfo{author}{\bibfnamefont{H.~H.} \bibnamefont{{Michels}}},
  \bibnamefont{and}
  \bibinfo{author}{\bibfnamefont{R.}~\bibnamefont{{C{\^o}t{\'e}}}},
  \bibinfo{journal}{Chem. Phys. Lett.} \textbf{\bibinfo{volume}{542}},
  \bibinfo{pages}{138} (\bibinfo{year}{2012}).

\end{thebibliography}

\end{document}